\documentclass[12pt,preprint]{aastex}    
\usepackage{lscape}
\usepackage{psfig}

\slugcomment{To appear in the May 10 2004 edition of the ApJ.}

\shortauthors{Strickland \etal}
\shorttitle{Quantifying supernova feedback}

\protect
\protect        
\protect        
\protect  

\newcommand{\eg}{{\rm e.g.\ }}

\newcommand{\ie}{{\it i.e.\ }}

\newcommand{\etal}{{\rm et al.\thinspace}}

\newcommand{\cm}{{\rm\thinspace cm}}
\newcommand{\km}{{\rm\thinspace km}}
\newcommand{\pcc}{\hbox{$\cm^{-3}\,$}}
\newcommand{\s}{{\rm\thinspace s}}
\newcommand{\yr}{{\rm\thinspace yr}}
\newcommand{\erg}{{\rm\thinspace erg}}

\newcommand{\pyr}{\hbox{\yr$^{-1}$}}
\newcommand{\ergps}{\hbox{$\erg\s^{-1}\,$}}

\newcommand{\kmps}{\hbox{$\km\s^{-1}\,$}}
\newcommand{\pcmsq}{\hbox{$\cm^{-2}\,$} }

\newcommand{\halpha}{H$\alpha$}

\newcommand{\K}{{\rm K}}
\newcommand{\hi}{H{\sc i}}

\newcommand{\nii}{[N{\sc ii}]}

\newcommand{\kpc}{{\rm\thinspace kpc}}
\newcommand{\Mpc}{{\rm\thinspace Mpc}}

\newcommand{\Lsol}{\hbox{$\thinspace L_{\sun}$}}
\newcommand{\Msol}{\hbox{$\thinspace M_{\sun}$}}
\newcommand{\Zsol}{\hbox{$\thinspace Z_{\sun}$}}

\begin{document}

\title{A high spatial resolution X-ray and H$\alpha$
study of hot gas in the halos of star-forming disk galaxies. 
II. Quantifying supernova feedback}

\author{David K. Strickland,\altaffilmark{1,2}
	Timothy M. Heckman,\altaffilmark{2}
	Edward J.M. Colbert,\altaffilmark{2}
	Charles G. Hoopes,\altaffilmark{2} and
	Kimberly A. Weaver.\altaffilmark{3}}

\altaffiltext{1}{{\it Chandra} Fellow.}
\altaffiltext{2}{Department of Physics and Astronomy, 
	The Johns Hopkins University,
	3400 North Charles Street, Baltimore, MD 21218}
\altaffiltext{3}{NASA/Goddard Space Flight Center, 
	Code 662, Greenbelt, Maryland 20771}

%

\begin{abstract}
We investigate how the empirical properties of hot X-ray-emitting gas
in a sample of seven starburst and three normal edge-on spiral galaxies
(a sample which covers the full range of star-formation intensity found
in disk galaxies) correlate with the size, mass, star formation rate
and star formation intensity in the host galaxies.
From this analysis we investigate various aspects of mechanical energy
``feedback'' --- the return of energy to the ISM 
from massive star supernovae and stellar winds --- on galactic scales.
The X-ray observations make use of the unprecedented
spatial resolution of the {\it Chandra} X-ray observatory to
remove X-ray emission from point sources more accurately than in any
previous study, and hence
obtain the X-ray properties of the diffuse thermal emission alone.
Intriguingly, the diffuse X-ray properties 
of the normal spirals (both in their disks
and halos) fall where extrapolation of the trends from the 
starburst galaxies with superwinds would predict.
We demonstrate, using a variety of multi-wavelength 
star formation rate and intensity indicators, that the luminosity
of diffuse X-ray
emission in the disk (and where detected, in the halo) is directly 
proportional to the rate of mechanical energy feedback from 
massive stars in the host galaxies. 
Accretion of gas from the IGM does not
appear to be a significant contributor to the diffuse X-ray emission 
in this sample. Nevertheless, with only 
three non-starburst normal spiral galaxies it is hard to exclude 
an accretion-based origin for extra-planar diffuse X-ray emission around
normal star-forming galaxies.
Larger galaxies tend to have more extended X-ray-emitting halos, but 
galaxy mass appears to play no role in determining
the properties of the disk or extra-planar X-ray emitting plasma.
 The combination of these luminosity and size correlations
leads to a correlation between the surface brightness of the diffuse
X-ray emission and  the mean star formation rate per unit area in the disk
(calculated from the Far-Infrared luminosity and the optical size of the
galaxy, $L_{\rm FIR}/D_{25}^{2}$).
Further observational work of this form will allow empirical constraints
to be made on the critical star formation rate per unit disk area 
necessary to blow hot gas out of the disk into the 
halo \citep{maclow88}. We show that
a minor generalization of standard superbubble theory directly
predicts a critical star formation rate per unit area for superbubble blow
out from the disk, and by extension for superwinds to blow out of the gaseous
halos of their host galaxy. 
At present there are a variety of poorly-known parameters
in this theory which complicate comparison between observation 
and theory, making it impossible to assess the quantitative accuracy
of standard superbubble blow out theory.
We argue that the crucial spatial region
around a galaxy that controls whether gas in starburst-driven superwinds
will escape into the IGM
is not the outer halo $\sim 100$ kpc from the host galaxy, but
the inner few halo scale heights, within $\sim 20$ kpc of the
galaxy plane. Given the properties of the
gaseous halos we observe, superwind
outflows from disk galaxies of mass $M \sim 10^{10}$ 
-- $10^{11} \Msol$ should still eject some fraction of their material
into the IGM.
\end{abstract}

\keywords{ISM: jets and outflows --- ISM: bubbles ---
galaxies: individual (NGC 253; NGC 891; NGC 1482; NGC 3034 (M82); 
	NGC 3073; NGC 3079; 
	NGC 3628; NGC 4244; NGC 4631; NGC 4945; NGC 6503) 
--- galaxies: halos --- galaxies: starburst --- X-rays: galaxies}

\section{Introduction}
\label{sec:introduction}

Massive stars exercise a profound influence over the baryonic component
of the Universe, through the return of ionizing radiation, and
via supernovae (SNe), kinetic energy and metal-enriched gas, back into the
interstellar medium (ISM) from which they form --- 
usually called ``feedback''. 
Feedback influences gas-phase conditions in the immediate
environment of the clusters within which the massive stars 
form \citep{mckee95,hollenbach97,wiseman98,pudritz00}, 
on galactic scales the phase structure and energetics of the ISM 
\citep{mckee77,cox81,norman89,norman96},
and on multi-Mpc scales the thermodynamics and enrichment of
the inter-galactic medium (IGM --- \eg see \citealt*{chiang88,ham90,shapiro94,voit96,heckman99,aguirre01}).

The vast range of spatial scales involved is only one of the difficulties 
encountered in attempting to study feedback. Restricting the discussion
to purely mechanical feedback from SNe and stellar winds (commonly 
termed SN feedback), a further 
difficulty is the broad range of
complicated gas-phase physics -- (magneto)hydrodynamic
effects such as shocks and turbulence, thermal conduction, and non-ionization
equilibrium emission processes. 
A final complication is that much of the energy and
metal-enriched material involved is in the hard-to-observe coronal 
($T \ga 10^{5}$ K) and hot ($T \ga 10^{6}$ K) gas phases, necessitating the
use of space-based FUV and X-ray telescopes.

X-ray observations of the halo-regions of nearby edge-on disk galaxies
provide the best single probe of the action of mechanical feedback 
on the galactic scale. Optical spectroscopy and imaging, 
sensitive to warm neutral and ionized gas with $T \sim 10^{4}$ K, 
are a less direct probe of mechanical
feedback as much of the excitation energy is supplied by ionizing radiation
from massive stars (radiative feedback), even when the kinematics
and spatial distribution
of this gas are controlled by SN feedback.
If mechanical feedback
processes transport mass, newly-synthesized metals, and energy from galaxies
into the IGM, then this material must pass through the halo of the host galaxy
on its way into the IGM, where its properties can be ascertained
before it expands and fades into observational invisibility.

Diffuse thermal X-ray emission is seen to extend to large
distances ($5$ to 20 kpc) above the planes of edge-on {\em starburst}
galaxies (e.g. \citealt*{fhk90,armus95,rps97,dwh98}),
which also show unambiguous optical emission line evidence for energetic
galactic scale outflows --- ``superwinds''
\citep{ham90}. These outflows are potentially very significant
in enriching and heating the IGM \citep{heckman99}. 

Both fully-numerical and semi-analytical
theoretical modeling
\citep{suchkov94,silich98,babul} suggest that the properties
of the gaseous halos these superwinds expand into play a major role in
determining the fate of the material in superwinds
(i.e. escape into the IGM, or confinement and eventual return to the disk).
Although simulations of starburst-driven winds should be treated
with caution, given the lack of calibration against observational
data, one robust result is the sensitivity of the wind properties
(in particular mass and metal loss to the IGM) on the
disk and halo ISM properties 
\citep[\eg][]{deyoung94,suchkov94,silich98,ss2000,
silich01}. In our view, the emphasis given to
galactic mass as a primary parameter influencing gas and metal ejection
efficiencies \citep[\eg][]{dekel86,maclow99,ferrara_tolstoy00,fps00}
blinds non-expert readers to the more fundamental role played by the
poorly-known disk and halo ISM distributions.
The gravitational potential only enters at a secondary level, by
influencing the gas distribution and finally via the escape
velocity.
Redressing the lack of observational knowledge about the 
gaseous halos that superwinds must pass through 
is one motivation for studying
the halos of \emph{normal} star-forming
galaxies.

Modern theories of the ISM in normal
spiral galaxies also predict hot gas in the halo, due to the
interchange of material between the disk and halo through 
``galactic fountains'' \citep{shapiro76,bregman80,avillez00},
or in a more-localized form through ``chimneys'' \citep{norman89}.
Energy feed back into the ISM from supernovae is believed to 
create bubbles of hot, metal-enriched,  gas, which
blow out of the disk and vent their contents into the halo of the
galaxy. After $\ga 10^{8}$ years or so, this
material cools and falls back to the disk,
possibly in a form analogous to the Galactic high velocity clouds.

Unfortunately, many of the fundamental aspects of these fountain/chimney
disk/halo interaction
theories are observationally unconstrained, fundamentally due to
our {\em current} lack of knowledge about the X-ray properties
of spiral galaxy halos. In the theoretical paradigm, fountains 
are primarily driven by the over-pressurized hot gas created by SNe, and 
the most direct diagnostic of the energetics, and elemental composition, of
this gas is the thermal X-ray emission it produces.

Studies of nearby edge-on star-forming galaxies 
with {\it Einstein} and the {\it ROSAT PSPC}
revealed {\em only one} normal spiral, NGC 891 \citep{bregman_and_pildis},
with X-ray-emitting gas in its halo\footnote{NGC 4631 is often
presented as an example of a
 ``normal'' spiral galaxy with a X-ray-emitting halo 
(\eg see \citealt{wang2001}). However, its
large-scale-height non-thermal radio halo, warm IRAS 60 to $100\mu$m flux
ratio, high \halpha~luminosity
and optically disturbed morphology show that it is a 
highly atypical spiral galaxy. 
It is a good example of a galaxy experiencing a mild, disk-wide,
starburst event \citep{golla94b}. We believe that the {\it ROSAT} PSPC-based 
detection of a $L_{X} = 2.6 \times 10^{39}$ diffuse X-ray halo around 
NGC 4565, claimed by \citet{vogler95}, is incorrect, as the archival
 2.9 ksec {\it Chandra} ACIS-S observation (ObsID 404)
reveals that the X-ray emission
$\sim 2\arcmin$ east of the nucleus (that comprised the majority of the 
emission originally considered to be diffuse) 
is due to two point sources.}. Although edge-on 
normal spiral galaxies did not receive as much attention with {\it Einstein}
\& {\it ROSAT} as other classes of galaxy, 
those observations that were performed
failed to produce believable detections of hot halo gas
in other edge-on\footnote{Diffuse X-ray emission in approximately face-on
star-forming galaxies (\eg \citealt{snowden95,cui96,ehle98}), 
can not be unambiguously associated 
with a hot halo, for obvious reasons.} normal 
star-forming galaxies 
\citep*{bregman82,vogler95,bregman97,fabbiano97,benson00}.
X-ray studies of our own galaxy's halo and disk/halo interface
(e.g. \citealt{kuntz00}) provide information complementary to the
study of other galaxies, but are difficult to perform
and interpret due to our location within the disk. 

More progress on investigating the disk/halo interface in normal
galaxies has been made at other wavelengths, primarily in the optical,
where extra-planar \halpha~emission 
\citep*[\eg][]{dettmar92,rand96,hoopes99,collins00,rossa03,miller03} 
and dust \citep{howk99}
has been detected in a variety of nearby, normal spiral galaxies.
These studies have shown that the presence of such
tracers of disk/halo interaction increases with higher levels
of star-formation intensity
in the underlying disk, but the mechanisms behind these
phenomena, and their energetics, are still unclear.

It is also not clear to what extent fountains are a distinct phenomenon
from superwinds. It is not unreasonable to think that physical
conditions and kinematics of  
the gas making up a fountain may be quite different from
that in a superwinds, despite the general similarity in driving mechanism.
For example, the filling factor of the dominant 
X-ray-emitting plasma might be high in a fountain \citep{bregman80}, 
as compared to superwinds where the filling factor is low 
\citep{chevclegg,suchkov94,ss2000,strickland00}. 
Another intriguingly hint of a difference comes from 
the {\it ROSAT} PSPC-based sample of both normal
spiral galaxies and starburst galaxies of all inclinations by 
\citet{read01}. They find that the X-ray-emitting gas ``in the coronal systems
[\ie galaxies assumed to have a hydrostatic hot gas corona/halo]
appears to be cooler than that seen in the wind systems [\ie starburst
galaxies with superwinds]''.

Practitioners of semi-analytical (e.g. \citealt{benson00}), and 
numerical cosmological models (e.g. \citealt{toft02}) of galaxy
formation and evolution offer a radically different 
origin for X-ray emitting 
halos around spiral galaxies.
Inter-galactic gas continually flows into galactic halos, is 
heated in an accretion shock 
to the halo virial temperature (although \citet{katz03} discuss
simulations in which only a small fraction of the accreted gas is heated to
halo virial temperatures), 
and then cools and accretes onto the disk. These models
predict red-shift zero spiral galaxy X-ray luminosities at, or 
just below, the {\it ROSAT}-based 
observational upper limits. The predicted
X-ray luminosity is a strong function of the galaxy mass, 
$L_{X} \propto v_{\rm rot}^{5}$,
and that the temperature of the gas is 
similar to the halo virial temperature $kT \sim 0.5 \, \mu m_{\rm H}
\, v_{\rm rot}^{2}$.
The majority of the diffuse X-ray emission arises close to the
disk, the X-ray surface brightness dropping off semi-exponentially
with height above the disk. These halo properties 
should be largely independent of the host disk's SF intensity, 
and the exact numerical implementation of SN feedback 
(J. Sommer-Larsen, 2002, private communication).

\clearpage
\begin{deluxetable}{llccrrrrrrrrrr}
\tabletypesize{\scriptsize}%
\rotate
\tablecolumns{14} 
\tablewidth{0pc} 
\tablecaption{Basic physical properties of the sample galaxies
	\label{tab:galaxies}} 
\tablehead{ 
\colhead{Galaxy} & \colhead{$\alpha, \delta$ (J2000)} 
	& \colhead{$i$}
	& \colhead{PA}
	& \colhead{$v_{\rm rot}$}
	& \colhead{D}
	& \colhead{scale}
	& \colhead{$f_{60}$}
        & \colhead{$f_{60}/f_{100}$}
	& \multicolumn{3}{c}{$L/(10^{10} \Lsol)$}
	& \colhead{$L_{H\alpha}$} 
	& \colhead{$M_{\rm TF}$}\\
\colhead{} & \colhead{(h m s, $\degr$ $\arcmin$ $\arcsec$)}
	& \colhead{($\degr$)}
	& \colhead{($\degr$)}
        & \colhead{(km/s)}
        & \colhead{(Mpc)}
        & \colhead{(pc)}
        & \colhead{(Jy)}
	& \colhead{}
	& \colhead{($L_{\rm IR}$)}
	& \colhead{($L_{\rm K}$)}
	& \colhead{($L_{\rm B}$)} 
	& \colhead{($10^{40}\ergps$)}
	& \colhead{($10^{10} \Msol$)}\\
\colhead{(1)} & \colhead{(2)}
	& \colhead{(3)}
	& \colhead{(4)}
	& \colhead{(5)} & \colhead{(6)}
	& \colhead{(7)} & \colhead{(8)} 
	& \colhead{(9)} & \colhead{(10)}
	& \colhead{(11)} & \colhead{(12)} & \colhead{(13)}
	& \colhead{(14)}
	}
\startdata

M82 & 09 55 51.9 \phm{-}+69 40 47.1$^{a}$
	& 73$^{b}$, 82$^{c}$ & 65$^{d}$
	& 137$^{a}$ & 3.6$^{e}$ & 17.5 & 1313.5$^{f}$ & 0.97 
	& 5.36 & 0.75 & 0.33 & 7.0$^{g}$ & 1.9 \\
NGC 1482 & 03 54 39.3 \phm{+}-20 30 08.9$^{h}$
	& 58$^{i}$ & 103$^{j}$ 
	& 165$^{k}$ & 22.1 & 107.1 & 35.3$^{f}$ & 0.77
	& 5.00 & 0.84 & 0.38 &  N 9.4$^{l}$ & 3.7 \\
NGC 253 & 00 47 33.2  \phm{+}-25 17 16.2$^{m}$ 
	& 79$^{n}$, 72$^{o}$ & 52$^{d}$, 49$^{o}$
	& 225$^{m}$ & 2.6$^{p}$ &  12.6 & 936.7$^{f}$ & 0.50 
	& 2.10 & 0.89 & 0.58 & 3.6$^{q}$ & 10.6 \\
NGC 3628 & 11 20 16.95 \phm{-}+13 35 20.1$^{r}$
	& 87$^{s}$, 80$^{r}$ & 104$^{d}$
	& 229$^{t}$ & 10.0$^{u}$ & 48.5 & 51.6$^{f}$ & 0.50
	& 1.74 & 1.58 & 1.06 & N 2.3$^{v}$ & 11.3 \\
NGC 3079 & 10 01 57.8 \phm{-}+55 40 47.2$^{w}$
	& 85$^{w}$ & 165$^{d}$
	& 244$^{t}$ & 17.1 & 82.9 &  50.2$^{f}$ & 0.49 
	& 4.76 & 1.55 & 0.98 & N 9.1$^{d}$ & 14.1 \\
NGC 4945 & 13 05 25.3 \phm{+}-49 29 09.0$^{x}$
	& 78$^{y}$ & 43$^{y}$ 
	& 184$^{t}$ & 3.7 & 17.9 & 588.1$^{z}$ & 0.42
	& 2.70 & 0.94 & 0.73 & N 2.4$^{d}$ & 5.2 \\
NGC 4631 & 12 42 07.2 \phm{-}+32 32 31.9$^{aa}$
	& 85$^{ab}$, 81$^{ac}$ & 86$^{ac}$, 83$^{ac}$
	& 150$^{ab}$ & 7.5$^{ab}$ & 36.4 & 82.9$^{z}$ & 0.40 
	& 1.74 & 0.62 & 0.98 & N 14.3$^{ad}$ & 2.6 \\
\tableline
NGC 6503 & 17 49 26.4 \phm{-}+70 08 39.7$^{ae}$
	& 75$^{af}$ & 123$^{af}$ 
	& 120$^{ag}$ & 5.2$^{ah}$ & 25.2 & 10.2$^{f}$ & 0.35 
	& 0.12 & 0.14 & 0.18 & 0.76$^{ai}$ & 1.2 \\
NGC 891 & 02 22 33.2 \phm{-}+42 20 56.2$^{aj}$
	& 89$^{aj}$ & 23$^{aj}$ 
	& 225$^{t}$ & 9.6 & 46.5 & 61.1$^{z}$ & 0.31
	& 2.47 & 1.66 & 0.78 & 3.3$^{ae}$ & 10.6 \\
NGC 4244 & 12 17 29.7 \phm{-}+37 48 20.4$^{ak}$
	& 85$^{ak}$ & 48$^{ak}$ 
	& 100$^{ak}$ & 3.6 & 17.5 & 4.2$^{z}$ & 0.26 
	& $\le 0.02$ & 0.04 & 0.08 & 0.42$^{ad}$ & 0.6 \\
\enddata 
\tablecomments{Column 2: Coordinates of the dynamical center of the
        galaxy, where available.
	Column 3: Inclination.
	Column 4: Position angle of the galactic disk. Multiple values
        for the inclination and position angle are given when
        several values are often reported in the literature.
	Column 5: Inclination-corrected circular velocity within the disk.
        Column 6: Assumed distance. Unless a specific reference is
	given, distances were calculated using the galaxy recessional
	velocity with respect to the microwave background \citep{rc3},
	and $H_{0} = 75 \kmps \Mpc^{-1}$.
	Column 7: Physical distance 
	corresponding to an angular size of 1 arcsecond.
	Column 8: IRAS $60\micron$ flux in Janskys.
	Column 9: IRAS $60$ to $100\micron$ flux ratio.
	Column 10: Total IR luminosity based on the observed IRAS fluxes, 
	$L_{IR} = 5.67\times10^{5} 
	D^{2}_{\rm Mpc} \times  (13.48 f_{12} + 5.16 f_{25} + 2.58 f_{60} + f_        {100}) \Lsol$, where $D_{\rm Mpc}$ is the distance to the galaxy
        in Mpc \citep{sander96}. 
	Given the large angular size of these local
        galaxies we use IRAS fluxes from extended source analyses,
        the exact source of the IRAS data used
        is given in a case by case basis above.
	Where multiple measurements
	were available for any galaxy, we used the following sources in 
	order of preference: \citet{rice88}, and \citet{soifer89}.
	Column 11: K-band luminosity, calculated from the K-band total
	magnitudes given in \citet{2mass_largegals} and converted to
	the Cousins-Glass-Johnson system using the transformations
	presented in \citet{carpenter01}.
	Zero points for conversion from magnitude to	
	flux were taken from \citet*{bessel98}.  Note that the units
	are $\Lsol$, not $L_{\rm K, \odot}$.
	Column 12: B-band luminosity from RC3 catalog $B_{T}$
	magnitude \citep{rc3}, and corrected for Galactic extinction
	using reddening values from \citet{burstein82}. 
	Note that no correction for extinction internal
	to the specific galaxy has been made.
	Column 13:  Observed \halpha-luminosity in units of 
	$10^{40} \ergps$. Note these values have {\em not} been
	corrected for extinction. Measurements
	that have not been corrected for \nii~emission are
	marked with a preceding $N$.
	Column 14: Baryonic mass (stellar plus \hi) 
	estimated from the K-band Tully Fisher relationship of \citet{bell01},
	where $M_{\rm TF} = 10^{9.79} \times (v_{\rm rot}/100 
	\kmps)^{3.51} \Msol$.}
\tablerefs{(a) \citet{weliachew84}, (b) \citet{ichikawa95}, 
	(c) \citet{lynds63}, (d) \citet{lehnert95}, (e) \citet{freedman94},
	(f) \citet{soifer89}, (g) \citet*{mccarthy87},
	(h) 2MASS Second Incremental Release,
	(i) \citet{hameed99}, 
	(k) Calculated assuming $v_{\rm rot} = W_{20}/2\sin^{2} i$, 
	using the $W_{20}$ value from \citet*{roth91},
	(l) \citet{hameed99},	
	(m) \citet{sorai2000}, (n) \citet{pence81},
	(o) \citet{puche91}, 
	(p) \citet{puche88}, 
	(q) From \halpha~images published in \citet{strickland02},
 	(r) \citet{douglas96},
	(s) \citet{irwin96}, (t) \citet{sofue97},  (u) \citet{soifer87}
	(v) \citet{fhk90}, 
	(w) \citet{irwin91}, 
	(x) \citet{ott2001},
	(y) \citet{koorneef93}, 
	(z) \citet{rice88},
	(aa) Dynamical center, \citet{golla94b}, (ab) \citet{rand94},
	(ac) \citet{golla99},  	(ad) \citet{hoopes99},
	(ae) Central optical \& X-ray 
	source, \citet*{lira02},	
	(af) \citet{rc3},  (ag) \citet{bottema97},
	(ah) \citet{karachentsev97}, 
	(ai) \citet{kennicutt98a}. 
	(aj) Nuclear 1.4 GHz continuum point source, \citet{rupen91},
	(ak) \hi~dynamical center, \citet{olling96}.
	}
\end{deluxetable} 
\clearpage
In order to investigate some of these questions we have compiled a
sample of 10 edge-on star-forming galaxies which have both
ground-based narrow-band optical imaging (probing extra-planar
gas at $T \sim 10^{4}$) and X-ray observations with the
{\it Chandra} X-ray Observatory. The spatial resolution
of the {\it Chandra} telescope ($\sim0\farcs8$ FWHM) 
is an order of magnitude better
than any previous or currently-existing X-ray instrument.
As approximately half of the X-ray luminosity
of a star-forming galaxies is due to X-ray luminous point 
sources, accurate studies of the X-ray diffuse emission
require high spatial resolution in order to allow point source
removal.

In \citet[henceforth referred to as Paper I]{strickland03} 
we described the sample, the X-ray and optical
data reduction performed, and presented results and 
interpretation of a spatial and spectral 
analysis of the diffuse X-ray emission.

In this paper we investigate how
the empirical properties of the hot gas, in particular
gas in the halos of these galaxies, 
correlate with the size, mass, star formation rate
and star formation intensity in the host galaxies.
From this analysis we investigate various aspects of mechanical energy
``feedback'' --- the return of energy to the ISM 
from massive star supernovae and stellar winds --- on galactic scales.

We describe the sample and their star-formation 
properties in \S~\ref{sec:sample}. In \S~\ref{sec:paper1_summary}
we provide a brief summary of the results from Paper I, before
moving onto considering what correlations exist between the
properties of the diffuse X-ray emission and other multi-wavelength
properties of the galaxies in \S~\ref{sec:discussion}. In 
\S~\ref{sec:discussion:theory} we discuss these results with respect
to various aspects of feedback-related theory, in particular the
conditions necessary for blow-out of gas from the disk and halo in
normal galaxies and in starbursts,
and the fraction of supernova mechanical energy available to drive
large-scale ISM motions. Our results and conclusions are summarized
in \S~\ref{sec:conclusions}.

\section{Sample galaxies}
\label{sec:sample}

The sample is formed of seven starburst galaxies and three  ``normal'' 
star-forming galaxies, all approximately edge-on ($i \ga 60\degr$)
moderate luminosity ($L_{\rm BOL} \sim 10^{10} \Lsol$), moderate-mass 
($M \sim 10^{10}$ -- $10^{11} \Msol$) disk galaxies within
a distance of $D \la 20 \Mpc$. This sample contains all the disk galaxies
for which {\it Chandra} ACIS data was available to us as of 2002 
September, including both public data and our own propriety observations.
Although the sample size is small, in particular with respect to non-starburst
systems, it does span the full range of star formation activity
found in disk galaxies. The starburst galaxies in our sample
are all good examples of the ``typical'' starburst galaxy
in the local universe, \ie infrared warm galaxies with  a far-infrared
luminosity $L_{\rm FIR} \approx$ a few $\times 10^{10} \Lsol$ \citep{soifer87}.
The three non-starburst systems in the sample are NGC 891, NGC 6503 
and NGC 4244. NGC 891 is a well-studied spiral galaxy, believed 
to be similar to our own galaxy in mass and star formation rate. 
It is the {\em only} ``normal'' spiral galaxy (apart from possibly our own)
for which robust evidence of
extra-planar diffuse X-ray emission has been discovered
\citep{bregman_and_pildis}. Both NGC 6503 and NGC 4244 are relatively
low mass spiral galaxies. We refer the 
reader to Paper I for a more detailed discussion of the sample and
the distinction between starburst and normal spiral galaxies.

Basic physical properties of the
sample galaxies are shown in Table~\ref{tab:galaxies}. We distinguish
between starburst galaxies and more-normal star-forming galaxies on
the basis of their IR-warmth --- specifically the IRAS $60$ to $100\micron$
flux ratio $f_{60}/f_{100}$. Higher star formation rates per unit disk area
lead to higher dust temperatures, and hence higher $f_{60}/f_{100}$
ratios, with starburst galaxies typically having $f_{60}/f_{100} \ge 0.4$.
We will refer to those galaxies with $f_{60}/f_{100} < 0.4$ as the normal
spirals.

Intuitively, we would expect the effect of mechanical energy feedback
on the galactic scale to depend on the total star formation rate, and/or
the star formation rate per unit volume or area. Star formation and
supernova rates, and rates per unit area, for the sample galaxies are given in 
Table~\ref{tab:galactic_sf}. Note that galaxies are listed 
in all tables in order of decreasing $f_{60}/f_{100}$ ratio, \ie
approximately in order of decreasing star formation rate per unit area.

\subsection{Star formation rates and intensities}
We also consider two other simple, and widely-available 
proxies of the star-formation 
intensity: FIR and non-thermal radio face-on surface brightnesses,
quantified for our sample in Table~\ref{tab:galactic_sf}.
Although somewhat simplistic estimators of SF intensity, 
these do have the advantages of being available for 
almost all local disk galaxies from
uniformly-obtained datasets, 
and are distance-independent.

Very detailed studies of the star-formation rates,
spatial distributions, and history, are available for
the brightest starbursts 
such as M82 (\citealt*{rieke93,mcleod93,satyapal97,degrijs01,forster01}) 
and NGC 253 \citep{engelbracht98}, but measurements of similar quality 
are not available for many of the starburst
galaxies in our sample, let alone the normal spiral galaxies.
We did calculate star formation intensities for all our sample
galaxies based on all available optical, NIR and radio measurements
we could find in the literature, but found the scatter in estimated
SF intensity for any particular galaxy to be intolerably large.
\clearpage
\begin{deluxetable}{lrrrrrrrrrrrrrrrrrr}
 \tabletypesize{\tiny}%
\rotate
\setlength{\tabcolsep}{0.03in}
\tablecolumns{19} 
\tablewidth{0pc} 
\tablecaption{Star-formation and supernova rates, SF intensity, 
	and proxies thereof \label{tab:galactic_sf}} 
\tablehead{ 
\colhead{Galaxy} 
	& \colhead{SFR$_{H\alpha, T}$} 
	& \colhead{SFR$_{IR, T}$} 
	& \multicolumn{2}{c}{$D^{i}_{25}$}
	& \multicolumn{2}{c}{K-band $r_{0.5}$}
	& \colhead{$f_{\rm FIR}$}
	& \colhead{$f_{\rm FIR}/D_{25}^{2}$}
	& \colhead{$F_{\rm SN, FIR, D_{25}}$}
        & \colhead{${\cal R}_{SN, T}$} 
	& \colhead{$f_{\rm 1.4 GHz}$} & \colhead{$L_{\rm 1.4 GHz}$}
	& \multicolumn{2}{c}{$\theta_{\rm 1.4 GHz}$}
	& \colhead{$f_{\rm 1.4 GHz}/4\theta^{2}_{\rm 1.4 GHz}$} 
	& \colhead{$F_{\rm SN, FIR, \theta_{\rm 1.4 GHz}}$} 
	& \colhead{$\log \tau_{\rm gas}$}
	& \colhead{$\log \mu$} \\
\colhead{(1)} & \colhead{(2)}
	& \colhead{(3)}
	& \colhead{(4)}
	& \colhead{(5)} & \colhead{(6)}
	& \colhead{(7)} & \colhead{(8)} 
	& \colhead{(9)} & \colhead{(10)} 
	& \colhead{(11)} & \colhead{(12)}
	& \colhead{(13)} & \colhead{(14)}
	& \colhead{(15)} & \colhead{(16)}
	& \colhead{(17)} & \colhead{(18)} & \colhead{(19)}
	}
\startdata
M82   	& $>0.55$ & 9.2
	& 8.9 & 9.3 & 0.69 & 0.72 
	& 4745.9
	& 16.60 & 1211.0 & 0.107 
	& 7386.8 &  $5.47\times10^{4}$
	& 14.8 & 0.26 & 8.41 & $4.0\times10^{5}$  
	& 2.04 & 8.34 \\
NGC 1482  & $>0.74$ & 8.6
	& 2.2 & 14.1  & 0.10 & 0.66 
	& 136.8
	& 7.93 & 503.0 & 0.100 
	& 236.7 &  $7.91\times10^{4}$
	& 11.3 & 1.21 & 0.46 &  $1.7\times10^{4}$ 
	& 1.74 & 8.27 \\
NGC 253    & $>0.28$ & 3.6
	& 20.4 & 15.4  & 3.56 & 2.69
	& 4288.3
	& 2.86 & 177.1 & 0.042 
	& 5704.5 &  $1.38\times10^{4}$
	& 29.1 & 0.37 & 1.69 &  $7.8\times10^{4}$ 
	& 2.11 & 8.65 \\
NGC 3628   & $>0.18$ & 3.0
	& 10.5 & 30.5  & 1.67 & 4.86
	& 244.5
	& 0.606 & 37.5 & 0.035 
	& 470.2 & $1.98\times10^{4}$ 
	& 20.9 & 1.01 & 0.27 &  $8.5\times10^{3}$ 
	& 2.33 & 8.08 \\
NGC 3079   & $>0.72$ & 8.1
	& 5.5 & 27.4  & 0.71 & 3.54
	& 232.8
	& 2.14 & 127.1 & 0.129 
	& 820.7 & $1.53\times10^{5}$ 
	& 18.3 & 1.52 & 0.61 &  $1.4\times10^{4}$ 
	& 2.62 & 8.27 \\
NGC 4945   & $>0.19$ & 4.6
	& 13.8 & 14.9  & 2.88 & 3.09 
	& 2927.7
	& 4.28 & 246.5 & 0.054 
	& 4200.0 &  $3.92\times10^{4}$
	& 30.0 & 0.54 & 4.47 &  $1.9\times10^{5}$ 
	& 2.88 & 8.37 \\
NGC 4631   & $>1.13$ & 3.0
	& 10.5 & 22.9  & 1.51 & 3.29 
	& 412.3
	& 1.07 & 66.8 & 0.035 
	& 771.7 &  $1.71\times10^{4}$
	& 59.6 & 2.17 & 0.0543 &  $1.9\times10^{3}$ 
	& 2.48 & 7.70 \\
\tableline
NGC 6503   & $>0.06$ & 0.20
	& 5.6 & 8.4  & 0.77 & 1.18
	& 55.1
	& 0.492 & 16.7 & 0.0012 
	& 41.9 &  $6.74\times10^{2}$
	& 34.6 & 0.87 & 0.0088 &  $4.0\times10^{2}$ 
	& 2.97 & 8.23 \\
NGC 891   & $>0.26$ & 4.2
	& 9.3 & 26.1  & 1.47 & 4.10
	& 355.6
	& 1.14 & 39.5 & 0.049 
	& 285.7 &  $1.50\times10^{4}$
	& 55.2 & 2.57 & 0.0234 &  $1.9\times10^{3}$ 
	& 3.19 & 8.19 \\
NGC 4244   & $>0.03$ & 0.04
	& 10.2 & 10.7  & 1.85 & 1.94
	& 27.2
	& 0.071 & 3.5 & 0.0004 
	& 20.3 &  $5.65\times10^{1}$
	& 158.5 & 2.84 & 0.0003 &  $1.2\times10^{1}$ 
	&  4.53 & 7.72 \\
\enddata 
\tablecomments{Column 2: 
	Lower limits on the total galactic star formation rate 
	($\Msol \pyr$), based on the 
	\halpha~luminosity and using the formulae given in
	\citet{kennicutt98b}. Note that no correction for extinction
	has been applied.
        Column 3: Estimated total galactic star formation rate 
	($\Msol \pyr$), based on the IR luminosity, where
	SFR$_{IR} = 4.5 \times 10^{-44} L_{IR}$ [$\ergps$],
	again using the	\citet{kennicutt98b} formulae.
	Columns 4 and 5: Inclination-corrected 
	diameter of the stellar disk in arcminutes (4) and kpc (5), based 
	on the 
	$D_{25}$ values given in the RC3 and the distances given
	in Table~\ref{tab:galaxies}. The method of \citet{tully85}
	was used to correct the observed $D_{25}$ values for
	inclination.
	Columns 6 and 7: K-band half light radii in arcminutes (6) and
	kpc (7). Original values were taken from the 2MASS Large
	Galaxy Atlas \citep{2mass_largegals}.
	Column 8: IRAS Far-IR flux $f_{\rm FIR} = 2.58\times f_{60} + f_{100}$,
	in Jy. To convert to units of $\ergps \pcmsq$ multiply by
	$1.26\times10^{-11}$.
        Column 9: Far-IR luminosity per unit 
	stellar disk area $f_{\rm FIR}/D^{2}_{25}$,
	in distance-independent units of mJy arcsec$^{-2}$.
	To convert to the commonly used
	units of $10^{40} \ergps \kpc^{-1}$ 
	multiply $f_{\rm FIR}/D^{2}_{25}$ by 6.484.
	Column 10: One estimate of the mean 
	galactic supernovae rate per unit area,
	$F_{\rm SN, FIR, D_{25}} = {\cal R}_{SN}/D^{2}_{25}$, in units
	of SN Myr$^{-1}$ kpc$^{-2}$, is based on the star-formation rate
	calculated from $L_{\rm IR}$.
	Column 11: Estimated total galactic 
	core-collapse supernova rate ($\pyr$),
	where ${\cal R}_{SN} = 0.2 L_{IR}/10^{11} \Lsol$ \citep{ham90}.
	Column 12: 1.4 GHz radio flux in mJy, based on the NVSS data. See
	\S~\ref{sec:data_analysis:radio} for more details.
	Column 13: 1.4 GHz luminosity in units of Solar luminosities, 
	where $L_{\rm 1.4 GHz} 
	= 4\pi D^{2} \nu f_{\rm 1.4 GHz}$.
	The major-axis half-light radius, in units of arcseconds and kpc,
        that we derive the NVSS data is given in columns 
	14 and 15 respectively.
        Column 16: Effective 1.4 GHz surface brightness, in mJy arcsec$^{-2}$.
	Column 17: Another estimate of the mean SN rate per unit area
	$F_{\rm SN, FIR, \theta_{\rm 1.4 GHz}} = 
	{\cal R}_{SN}/4\theta^{2}_{\rm 1.4 GHz}$ (again
	in units of SN Myr$^{-1}$ kpc$^{-2}$), 
	based on the star-formation rate
	calculated from $L_{\rm IR}$ but using NVSS-based size $\theta_{1.4}$.
	Note that NGC 4945 was not observed in the NVSS. For NGC 4945 we
	use the nuclear region 1.4 GHz flux and size from \citet{elmouttie97},
	whose observations had similar spatial resolution to that of
	the NVSS.	
	Column 18: Logarithm of the gas consumption timescale in Myr,
	where $\log \tau_{gas} = \log \Sigma_{gas} - \log \Sigma_{\rm SFR}$,
	and $\Sigma_{gas}$ and $\Sigma_{\rm SFR}$ are the surface density
	of gas ($\Msol$ pc$^{-2}$) and star-formation rate ($\Msol$ 
	yr$^{-1}$ kpc$^{-2}$) respectively. Data are 
	from \citet{kennicutt98a}, except for NGC 1482,
	NGC 3628, NGC 4244, NGC 4631 
	and NGC 4945. For these galaxies the star formation rate was
	estimated from the IRAS fluxes. Gas masses were obtained from
	\citet{elfhag96} for NGC 1482 (using the CO brightness to $H_{2}$
	column density scaling in \citet{kennicutt98a}, and a
	radius of 2 kpc, as ISO observations by \citet{dale2000}
	demonstrate that 80\% of the 
	star-formation occurs within this radius), 
	\citet{irwin96} for NGC 3628 (central, $R \le 340$ pc, $H_{2}$ 
	gas mass), from \citet{olling96} for NGC 4244 (total H{\sc I}
	gas mass), from \citet{golla94b} for NGC 4631 ($H_{2}$ gas mass
	for $R \le 2.5$ kpc) and from \citet{dahlem93} for NGC 4945 
	(molecular gas mass within central 1.1 kpc).
	Column 19: Logarithm of the mean mass surface 
	density $\mu = M_{\rm TF}/D^{2}_{25}$.
	}
\end{deluxetable} 
\clearpage

The FIR luminosity divided by the optical isophotal diameter (in kpc)
at 25th magnitude, $L_{\rm FIR}/D^{2}_{25}$, is often used as a 
proxy for the mean star-formation rate per unit disk area in 
spiral galaxies (we shall occasionally refer to the
star formation rate per unit disk
area as the SFRI for convenience).
The presence of extra-planar \halpha~and/or radio emission,
and extra-planar dust, 
is known to qualitatively 
correlate with this parameter (\citealt{rand96}, see also
\citealt{dettmar98,howk99,rossa00,dahlem01}). It under-estimates
the true SF intensity in starburst galaxies, as the
star-formation is often much more concentrated than the old stellar light
measured by $D_{25}$. This method also over-estimates the mean SF
intensity in galaxies with star-formation rates somewhat less than that
of the Milky Way, where an increasing fraction of $L_{\rm FIR}$ is due
to cool dust heated by diffuse star-light, and not by massive stars.

Non-thermal radio emission, due to synchrotron emission from relativistic
electrons accelerated in supernova remnants, provides a good tracer
of SF activity on galactic scales \citep{condon92}. The 1.4 GHz luminosity
of a galaxy is almost directly proportional to the SF rate, which, combined
with the measured angular size of the radio-emitting region,
provides a good SF intensity measurement \citep{dahlem01}.

The radio fluxes, sizes and SF intensity values given in 
Table~\ref{tab:galactic_sf} are based on 1.4 GHz images
from the NRAO/VLA Sky Survey (NVSS, see \citealt{nvss}),
with the exception of southern galaxy NGC 4945, which
lies outside the region of sky surveyed by the NVSS.
A description of how these values were obtained is given in
\S~\ref{sec:data_analysis:radio}

\subsection{Radio fluxes, sizes and SF intensity measurements}
\label{sec:data_analysis:radio}

Radio fluxes and half-light radii were measured from the processed
NVSS 1.4 GHz radio images (\url{http://www.cv.nrao.edu/nvss/postage.shtml}).
Fluxes were summed within rectangular apertures encompassing
the optical disk of each galaxy, each of total
width equal to the optical $D_{25}$ diameter and breadth
equal to the minimum of 1\farcm5 (twice the angular 
resolution of the NVSS) or the angular equivalent of 4 kpc, and
centered on the nuclear positions given in Table~\ref{tab:galaxies}.
The major-axis half-light radii $\theta_{\rm 1.4}$ 
were measured from the radio flux within
the disk regions defined above using $\Delta r = 5\arcsec$-wide pixels. 
Inspection of the resulting radio surface brightness slices
demonstrated that the fluxes at the extreme ends of the slices were
in all cases statistically consistent with zero, allowing us to define
the 100\%-enclosed light radii as the edges of the surface brightness
profiles. Given the linear, rather than radial, nature of these
major-axis surface brightness profiles we obtained 
two half-light radii on either side of the nucleus for each galaxy, 
$\theta_{\rm +r}$  and $\theta_{\rm -r}$, which were combined to give 
the final half light radius $\theta_{\rm 1.4} = \{(\theta_{\rm +r}^{2}+
\theta_{\rm -r}^{2})/2\}^{1/2}$shown in Table~\ref{tab:galactic_sf}.

We did not use the fluxes and sizes given in the 
NVSS Source Catalog itself, which quotes fluxes and sizes from a
a 2-dimensional Gaussian model fit to the radio images.
The NVSS Source Catalog underestimates the total 1.4 GHz
flux from galaxies such as NGC 253, NGC 3628 and NGC 4631 by
a factor $\sim 2$,
where a relatively bright compact nuclear region of radio emission
is surrounded by a fainter extended radio-emitting disk, \ie
cases where a single Gaussian fit the the flux distribution does not
work well.

NGC 4945 is a southern hemisphere target and hence was not observed 
in the NVSS. For NGC 4945 we
use the nuclear region 1.4 GHz flux and size quoted in
\citet{elmouttie97}. Their observations had a 
similar spatial resolution to that of
the NVSS, making these values a reasonably good match to the values
we use for the other galaxies.

Note that the moderately-low spatial resolution 
of this radio data (FWHM = $45\arcsec$) will
over-estimate the spatial size of the radio emission
for the more compact starbursts (and hence under-estimate the
SF intensity), especially for the more distant
galaxies such as NGC 3079 and NGC 1482.

\clearpage
\begin{figure*}[!ht]
\epsscale{0.9}
\plotone{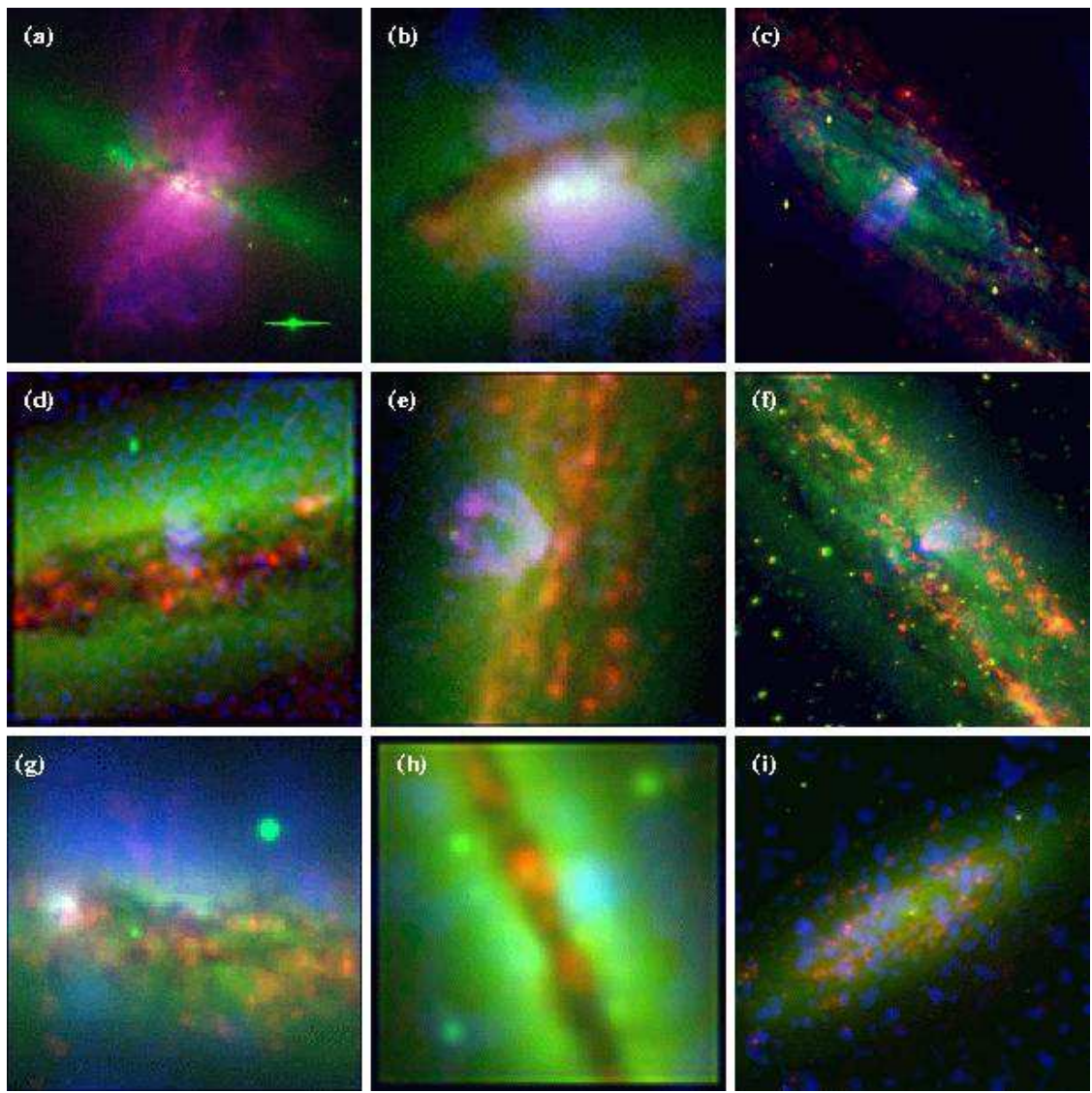}
\caption{False color composite images of the central 5 kpc by 5 kpc
of M82 (a), NGC 1482 (b), NGC 253 (c), NGC 3628 (d), NGC 3079 (e), 
NGC 4945 (f), NGC 4631 (g), NGC 891 (h) and NGC 6503 (i). \halpha~emission
is shown in red, optical R-band in green and 0.3--2.0 keV energy
band soft X-ray emission is shown in blue. The X-ray emission is the
point-source subtracted diffuse emission which has been adaptively
smoothed to achieve a local S/N of 3, which tends to over-smooth
structure in the X-ray emission. All images are shown on a square-root
intensity scale. In contrast to Figs.~2 
to 13 presented in Paper I, the same
absolute intensity is not used, in order to show some of the
fainter emission features. The galaxies are shown in order of
(approximately) decreasing star formation rate per unit area. Note
the close spatial similarities between the minor-axis-oriented
\halpha~emission and the diffuse X-ray emission, in particular 
the limb-brightened
outflow cones in NGC 1482, NGC 253, NGC 3079 and NGC 4945.}
\label{fig:rgb_circ}
\end{figure*}

\clearpage
\begin{figure*}[!ht]
\epsscale{0.9}
\plotone{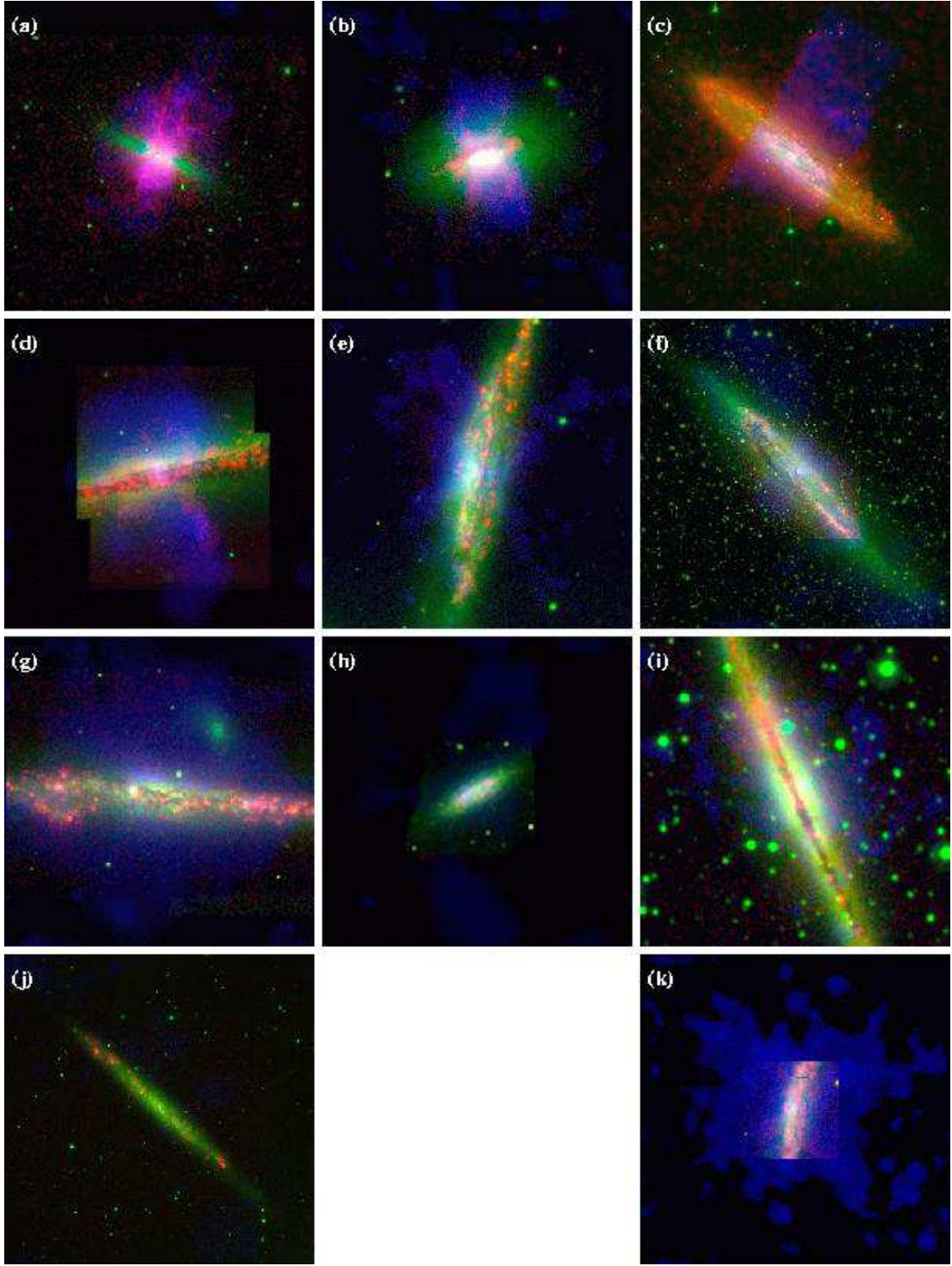}
\caption{As Fig.~\ref{fig:rgb_circ}, except that panels (a) through (j)
  now show 20 kpc by 20 kpc boxes, while in panel (k) the region is
  40 kpc by 40 kpc. The galaxies are M82 (a), NGC 1482 (b), NGC 253 (c),
  NGC 3628 (d), NGC 3079 (e and k), NGC 4945 (f), NGC 4631 (g), NGC 6503
  (h), NGC 891 (i), NGC 6503 (j), and NGC 3079 (k).}
\label{fig:rgb_halo}
\end{figure*}
\clearpage

\section{Summary of results from Paper I}
\label{sec:paper1_summary}

In Paper I \citep{strickland03} we presented 
a detailed study of the diffuse X-ray emission in
a sample of 10 approximately edge-on disk galaxies 
(7 starburst galaxies, 3 normal spirals) that span the full range of
star formation found in disk galaxies. With the arcsecond spatial
resolution of the {\it Chandra} X-ray telescope we were able to
separate the X-ray emission from point sources from the truly diffuse
component. The high spatial resolution also allowed 
a meaningful comparison to be made of the spatial location of the 
X-ray emission with respect to multi-wavelength data (in particular
to ground or space-based optical imaging). In addition to a detailed
discussion of each galaxy, we presented a mini-atlas of
soft and hard X-ray, \halpha~and R-band images of each of the 10 galaxies,
shown at a common spatial and surface brightness scale to facilitate
cross-comparison. Three color composite images of the diffuse soft X-ray
emission (blue) with optical \halpha~emission (red) and R-band stellar
continuum emission (green) are shown in 
Figs.~\ref{fig:rgb_circ} and \ref{fig:rgb_halo}.

We presented a variety of quantitative measures of the spatial extent, 
spectral hardness, and shape of the diffuse X-ray emission,
several of which can be applied to future samples
of galaxies with lower S/N data. We found that the spectral
properties of the diffuse X-ray emission show only weak, or some cases no, 
variation with increasing height (outside a heavily-absorbed, spectrally-harder
region within $z\sim2$ kpc of the disk mid-plane). 
The vertical decrease in surface brightness
of the extra-planar emission ($|z| \ge 2$ kpc) appears to better
described by exponential (effective surface brightness scale heights
are typically between 2 -- 4 kpc),
or Gaussian models, than the power law expected
of a freely expanding fluid.

For the eight galaxies with detections of extra-planar (\ie halo-region)
diffuse emission,
we find that a common spectral model, comprising a two-temperature
{\sc Mekal} \citep{mekal}
hot plasma model with an enhanced $\alpha$-to-Fe 
element ratio, can simultaneously fit the ACIS X-ray halo-region
spectra. The X-ray-derived metal abundances show super-Solar
ratios of $\alpha$-process elements (primarily oxygen) to
iron. This is consistent
with the origin of the X-ray emission being either
(metal-enriched) merged SN ejecta, 
or from shocked ambient halo ISM (with 
moderate depletion of refractory elements onto dust).

Although less luminous, the spatial and spectral properties of
the thermal X-ray emission in the normal, non-starbursting, spiral
galaxy NGC 891 are very similar to those of the starburst galaxies 
with superwinds.

Our favored model for the origin of the extra-planar soft X-ray emission 
is that SN feedback in the disks
of star-forming galaxies create, via blow out and venting of hot
gas from the disk, tenuous exponential atmospheres
of density scale height $H_{\rm g} \sim 4$ -- 8 kpc.
AGN-driven winds do not
appear to be significant in this sample, in that there is no
obvious correlation between the presence and luminosity of any AGN
and the properties of the diffuse X-ray emission.
The soft thermal X-ray emission observed in the halos of the starburst
galaxies is either pre-existing halo medium
(which has been swept-up and shock-heated by the starburst-driven superwinds)
or from a small fraction (by volume)
of the merged-SN-ejecta that has been compressed
near the walls of the outflow (\eg by a reverse shock propagating back 
into the outflowing wind).
In either case the observed
exponential X-ray surface brightness distributions are an inheritance
from galactic fountain
activity prior to the currently-observed starburst phase.
This model is based on the qualitative 
2-D morphology of the diffuse X-ray and optical
\halpha~emission (in particular the filamentary, occasionally limb-brightened
morphology of both the X-ray and \halpha~emission), 
as well as interpretation of the more-quantitative
minor axis surface brightness and spectral
hardness profiles. An implication of this model
is that galactic-scale gaseous halos may be common around 
star-forming disk galaxies. Observing starburst galaxies, in which superwinds
``light-up'' these pre-existing halos, may present the best method
for studying the gaseous halos of star-forming galaxies.
 
\clearpage
\begin{deluxetable}{llll}
  \tabletypesize{\tiny}%
  \rotate
  \tablecolumns{4}
  \tablewidth{0pc}
  \tablecaption{Summary and location of data values used in this paper 
	\label{tab:data_location}}
  \tablehead{
	\colhead{Value} & \colhead{Tabulated in} 
	& \colhead{Description} & \colhead{Appears in}  \\
	\colhead{(1)} & \colhead{(2)} &  \colhead{(3)} &  \colhead{(4)}
}  
\startdata
\cutinhead{Data values tabulated in this paper.}
$L_{\rm IR}$ 
	& Table~\ref{tab:galaxies} 
	& Total galactic IR luminosity, derived from {\it IRAS} fluxes. 
	& Figs.~\ref{fig:flux_flux1} and \ref{fig:lum_lum3} \\
$L_{\rm K}$ 
	& Table~\ref{tab:galaxies} 
	& Total galactic K-band luminosity, derived 
	from fluxes in the 2MASS Large Galaxy Atlas. 
	& Figs.~\ref{fig:flux_flux1} and \ref{fig:lum_lum3} \\
$L_{\rm B}$ 
	& Table~\ref{tab:galaxies} 
	& Total galactic B-band luminosity. Not corrected for extinction. 
	& Figs.~\ref{fig:flux_flux1} and \ref{fig:lum_lum3}\\
$v_{\rm rot}$
	& Table~\ref{tab:galaxies}
	& Inclination-corrected peak circular velocity.
	& Figs.~\ref{fig:halo_fluxratio}, \ref{fig:sizes_sizes1} 
	and \ref{fig:surfb_sfintens} \\
$f_{60}/f_{100}$
	&  Table~\ref{tab:galaxies}
	& IRAS 60 to $100\micron$ flux ratio, an indicator of 
	the intensity of star formation.
	&  Figs.~\ref{fig:halo_fluxratio} and \ref{fig:surfb_sfintens} \\
$M_{\rm TF}$
	& Table~\ref{tab:galaxies}
	& Galactic baryonic mass derived from the K-band Tully-Fisher relation.
	& Figs.~\ref{fig:lum_lum3} and \ref{fig:observed_fsn} \\
$L_{\rm 1.4 GHz}$ 
	& Table~\ref{tab:galactic_sf} 
	& Total galactic radio luminosity at 1.4 GHz, 
	derived from the {\it NVSS}. 
	& Figs.~\ref{fig:flux_flux1} and \ref{fig:lum_lum3} \\
$f_{\rm 1.4 GHz}$ 
	& Table~\ref{tab:galactic_sf} 
	& Total galactic radio flux at 1.4 GHz, 
	derived from the {\it NVSS}. 
	& \nodata \\
$\theta_{\rm 1.4 GHz}$
	& Table~\ref{tab:galactic_sf}
	& Radio major-axis half light radius.
	& Fig.~\ref{fig:sizes_sizes1} \\
$f_{\rm 1.4 GHz}/\theta^{2}_{\rm 1.4 GHz}$ 
	& Table~\ref{tab:galactic_sf} 
	& Radio-based estimate of the star formation rate intensity.
	& Figs.~\ref{fig:halo_fluxratio} and \ref{fig:surfb_sfintens} \\
$f_{\rm FIR}$
	& Table~\ref{tab:galactic_sf}
	& Total galactic IRAS FIR flux.
	& \nodata \\
$D_{25}$
	& Table~\ref{tab:galactic_sf}
	& Inclination corrected optical diameter at a surface brightness of 
	25 magnitudes per square arcsecond.
	& Fig.~\ref{fig:sizes_sizes1} \\
$f_{\rm FIR}/D^{2}_{25}$
	& Table~\ref{tab:galactic_sf}
	& Commonly used proxy of the mean galactic
	star formation rate per unit area.
	& Figs.~\ref{fig:halo_fluxratio} and \ref{fig:surfb_sfintens} \\
K-band $r_{0.5}$
	& Table~\ref{tab:galactic_sf}
	& K-band half light radius, from the 2MASS Large Galaxy Atlas. 
	& Fig.~\ref{fig:sizes_sizes1} \\
$\mu$
	& Table~\ref{tab:galactic_sf}
	& Mean galactic mass surface density.
	& Fig.~\ref{fig:sizes_sizes1} \\
${\cal R}_{\rm SN}$
	& Table~\ref{tab:galactic_sf}
	& Total galactic core-collapse supernovae rate, based on
	the total IR luminosity.
	& \nodata \\
$F_{\rm SN, FIR, D_{25}}$
	& Table~\ref{tab:galactic_sf}
	& One estimate of the mean supernovae rate 
	per unit disk area, where $F_{\rm SN, FIR, D_{25}} = 
	{\cal R}_{\rm SN}/D^{2}_{25}$  
	& Fig.~\ref{fig:observed_fsn} \\
$F_{\rm SN, FIR, \theta_{\rm 1.4 GHz}}$
	& Table~\ref{tab:galactic_sf}
	& Another estimate of the mean supernovae rate 
	per unit disk area, where $F_{\rm SN, FIR, D_{25}} = 
	{\cal R}_{\rm SN}/ 4 \theta^{2}_{\rm 1.4 GHz}$  
	& Fig.~\ref{fig:observed_fsn} \\
\cutinhead{Data values tabulated in Paper I.}
$\Sigma_{0.5}$
	& Table~4
	& Mean diffuse X-ray surface brightness between $z=0$ and the minor
	half light height $z_{0.5}$ and between $-5 \le r (kpc) \le 5$.
	& Fig.~\ref{fig:surfb_sfintens} \\
$z_{1e-9}$
	& Table~4
	& Height along the minor axis at which the diffuse X-ray surface
	brightness reaches $10^{-9}$ photons s$^{-1}$ cm$^{-2}$ arcsec$^{-2}$. 
	& Fig.~\ref{fig:observed_fsn} \\
$z_{0.95}$
	& Table~4
	& Minor axis diffuse X-ray 95\%-flux enclosed height.
	& Fig.~\ref{fig:observed_fsn} \\
$r_{0.5}$
	& Table~5
	& Major-axis diffuse X-ray half light radius.
	& Fig.~\ref{fig:sizes_sizes1} \\
$r_{0.75}$
	& Table~5
	& Major-axis diffuse X-ray 75\%-flux enclosed radius.
	& Fig.~\ref{fig:sizes_sizes1} \\
$H_{\rm eff}$
	& Table~6
	& Minor-axis diffuse X-ray exponential scale height 
	in the 0.3--1.0 keV energy band.
	& Figs.~\ref{fig:sizes_sizes1} and \ref{fig:observed_fsn} \\
$<kT>$
	& Table~8
	& X-ray luminosity weighted mean temperature.
	& Fig.~\ref{fig:sizes_sizes1} \\
$\Sigma_{\rm HALO}$
	& Table~8
	& Mean diffuse X-ray surface brightness within the halo spectral
	extraction region.
	& Fig.~\ref{fig:surfb_sfintens} \\
$f_{\rm X, HALO}$ 	
	& Table~9 
	& Halo region absorption-corrected X-ray flux (0.3--2.0 keV
	energy band).
	& Fig.~\ref{fig:halo_fluxratio} \\
$L_{\rm X, HALO}$ 
	& Table~9 
	& Halo region absorption-corrected X-ray luminosity (0.3--2.0 keV
	energy band).
	& Figs.~\ref{fig:flux_flux1} and \ref{fig:lum_lum3} \\
$L_{\rm X, DISK}$ 
	& Table~9 
	& Disk region absorption-corrected X-ray luminosity (0.3--2.0 keV
	energy band). 
	& \nodata \\
$L_{\rm X, NUCL}$ 
	& Table~9 
	& Nuclear region absorption-corrected X-ray luminosity (0.3--2.0 keV
	energy band). 
	& \nodata \\
$L_{\rm X, D+N}$ 
	& \nodata 
	& Sum of $L_{\rm X, NUCL}$ and $L_{\rm X, DISK}$. 
	& Figs.~\ref{fig:flux_flux1} and \ref{fig:lum_lum3} \\
\enddata
\tablecomments{Column 2: Table in this paper, or paper I, in which
	the symbol in Column 1 is tabulated. Column 3: A short description
	of the mean of each symbol -- see the appropriate Table for
	a more detailed explanation. Column 4: Figures in this
	paper in which the values are used. Some data values used in
	the figures, \eg $L_{\rm X, D+N}$, are not
	explicitly tabulated but can calculated from data that is
	tabulated.
}  
\end{deluxetable}
\clearpage

\section{Correlation analysis}
\label{sec:discussion}

We shall now move onto a more-quantitative comparison of the properties
of the diffuse X-ray with other properties of host galaxies.
For convenience, definitions of the data values we use 
(from both Paper I and this paper), the table numbers from 
which they were extracted, and the
figures in which they are used, are given in Table~\ref{tab:data_location}.
In Paper I we considered the thermal emission from the
nuclear region and the disk exterior to the nucleus separately. In this
paper we will combine the X-ray luminosities of the nucleus ($L_{\rm X,NUCL}$)
and the disk ($L_{\rm X,DISK}$) to give a total 
X-ray luminosity $L_{\rm X,D+N}$
from hot gas with $|z| < 2$ kpc of the plane of the galaxy. For convenience
we shall refer to this combination of Paper I's disk and nuclear
regions as the disk in the rest of this paper.
Halo region 
(\ie $|z| > 2$ kpc) luminosities are used as published in Paper I.

\begin{figure*}[!t]
\epsscale{0.9}
\plotone{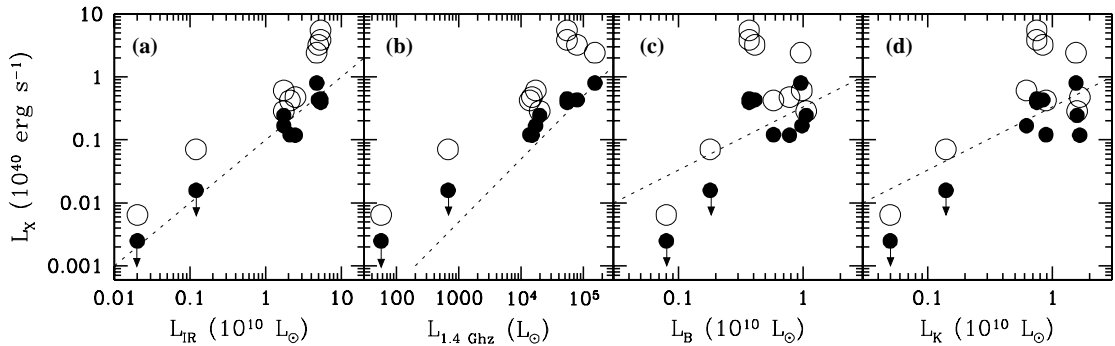}
\caption{Disk (including the nuclear region) and halo region
        diffuse emission 0.3--2.0 keV energy band
	X-ray luminosities ($L_{\rm X,D+N}$ 
        and $L_{\rm X,HALO}$, open and filled circles respectively), 
	plotted against the host galaxies' total IRAS IR luminosity 
	(panel a), the 1.4 GHz luminosity (panel b), B-band optical 
	luminosity (panel c) and NIR K-band luminosity (panel d).
	The dotted
	lines show the trend expected for a relationship of unit slope ---
	they are \emph{not} fits to the data. See \S~\ref{sec:discussion}
	for a definition of what is meant by disk and halo regions.}
\label{fig:flux_flux1}
\end{figure*}

Given the very significant absorption of
the diffuse emission from NGC 4945 by the high Galactic foreground hydrogen,
the low detection significance and atypical 
spectral hardness of the halo diffuse emission
(see Paper I), 
we have excluded it from further analysis. In particular, it is excluded
from Figs.~\ref{fig:flux_flux1} to Fig.~\ref{fig:observed_fsn}.

\subsection{Star-formation or galaxy mass? The physical-drivers of
	 extra-planar X-ray emission}
\label{sec:discussion:mass_or_sf}

The only previous study quantitatively comparing the \emph{diffuse}
X-ray emission\footnote{Significant 
effort has been expended in the past comparing the
multi-wavelength luminosities of galaxies in an attempt to elucidate
the physical causes of the \emph{integrated} X-ray emission 
in these systems. We refer the interested reader to 
\citet{fabbiano89}, \citet*{shapley01} and \citet{fabbiano02}.} 
from late-type, star-forming galaxies with
their integrated luminosities at other wavelengths is the {\it ROSAT} 
PSPC-based study published in  \citet{rps97} and \citet{read01}.
They find that the estimated
total diffuse emission X-ray luminosity per unit galaxy mass 
(estimated from $L_{\rm X}/L_{\rm B}$) correlates well with $L_{\rm B}$, 
an estimator of galaxy mass for \emph{normal} spirals. For starburst galaxies
$L_{\rm X}/L_{\rm B}$ shows a better correlation with $L_{\rm FIR}$,
\ie with a proxy of star formation rate.

There are a number of reasons while it is worthwhile to re-examine this issue,
apart from the obvious advantage of using the higher quality data 
from an instrument with greater sensitivity, 
and vastly superior capabilities of separating point-source and diffuse
emission.
Determining the properties of extra-planar hot gas in
normal galaxies has assumed a new significance, given its importance as
a diagnostic of the accuracy of theories of galaxy formation and evolution.
The normal galaxy sample of \citet{read01} is largely a collection of
face-on systems, with the approximately Milky-Way-like galaxy NGC 891 
\citep{kruit84,wainscoat87} being classed as a starburst on the
basis of the reddening-correction-sensitive
$L_{\rm FIR} > 0.38 \, L_{\rm B}$ criterion used by
\citeauthor{read01} 
as the definition of a starburst. It is only in high inclination
systems where we can separate the disk and halo diffuse components,
the latter being the most interesting component in testing both SN feedback
models and cosmological accretion models. 

There are only three approximately edge-on normal spiral 
galaxies in our sample, the Milky Way-like galaxy NGC 891 and
the two lower mass and lower luminosity spirals NGC 6503 and NGC 4244.
These three galaxies are too small a sample on their own
to establish any hard conclusions about the
properties of diffuse X-ray emission in normal spiral galaxies, 
but in combination with the starbursts we can make a meaningful
first assessment of how closely the
properties of the diffuse X-ray emission in these systems resemble those of
the edge-on starburst galaxies. Of course more
X-ray observational study of edge-on spiral galaxies is required before we
understand extra-planar hot gas in  normal spiral galaxies as well as
we understand the phenomenon in starbursts.

\subsection{X-ray and multi-wavelength luminosities}
\label{sec:discussion:mass_or_sf:rates}

We find reasonable correlations between the soft X-ray luminosity
of both the disk and halo diffuse X-ray emission, 
and the total galactic FIR and 1.4 GHz radio luminosities 
(see Fig.~\ref{fig:flux_flux1}), 
which are well-known measures of the total star-formation rate. 
In particular, note that the slope of the correlations between
$L_{\rm IR}$ or $L_{\rm 1.4 GHz}$ 
and $L_{\rm X}$ is of order unity, as expected if the
diffuse X-ray emission is caused by star-formation activity.

Correlation between the X-ray luminosities and the 
optical B and NIR K-band luminosities (\ie proxies for total
galactic stellar mass) is somewhat weaker,
in particular in the case of the halo X-ray luminosity when only
detections are considered.

\begin{figure*}[!t]
\epsscale{0.8}
\plotone{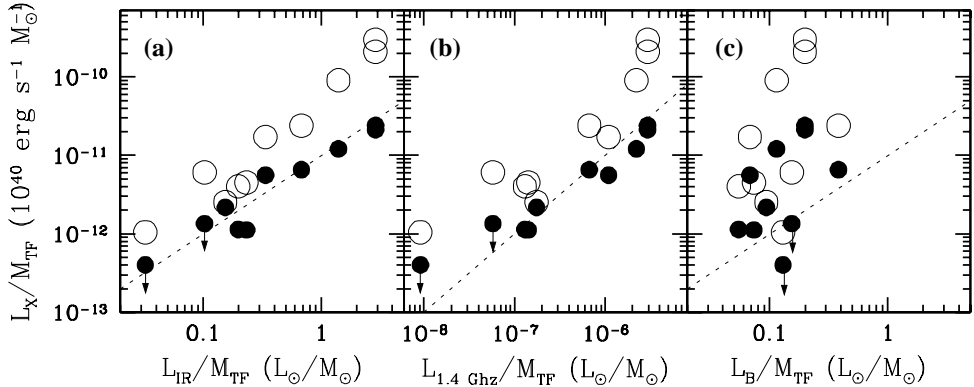}
\caption{As Fig.~\ref{fig:flux_flux1}, except all panels are luminosity
	per unit stellar mass. The disk and halo diffuse 
	emission 0.3--2.0 keV energy band X-ray luminosities
	divided by total galactic stellar mass 
	are shown
	and open and filled circles respectively. These values 
	are plotted against IR luminosity per unit mass (panel a), 
	radio luminosity per unit mass (panel b) and 
	B-band optical luminosity per unit mass (panel c). The dotted
	lines show the trend expected for a relationship of unit slope.}
\label{fig:lum_lum3}
\end{figure*}

In Fig.~\ref{fig:lum_lum3} we plot all luminosities normalized by the
stellar mass of the host galaxy. Estimates of the stellar mass of each galaxy
are based on the 2MASS K-band Tully-Fisher relationship
derived by \citealt{bell01}), and are listed in Table~\ref{tab:galaxies}.
As the K-band luminosity $L_{\rm K}$ and $M_{\rm TF}$ are not independent
variables we have not included a plot of $L_{\rm X}/M_{\rm TF}$ against 
$L_{\rm K}/M_{\rm TF}$ in this figure. Having taken out any dependence
related to galaxy mass (\ie more massive galaxies having more star-formation
and more X-ray emission than small galaxies), Fig.~\ref{fig:lum_lum3}
makes it clear that the diffuse X-ray luminosity per unit mass is directly
proportional to star-formation rate per unit mass.

\begin{figure*}[!ht]
\epsscale{0.9}
\plotone{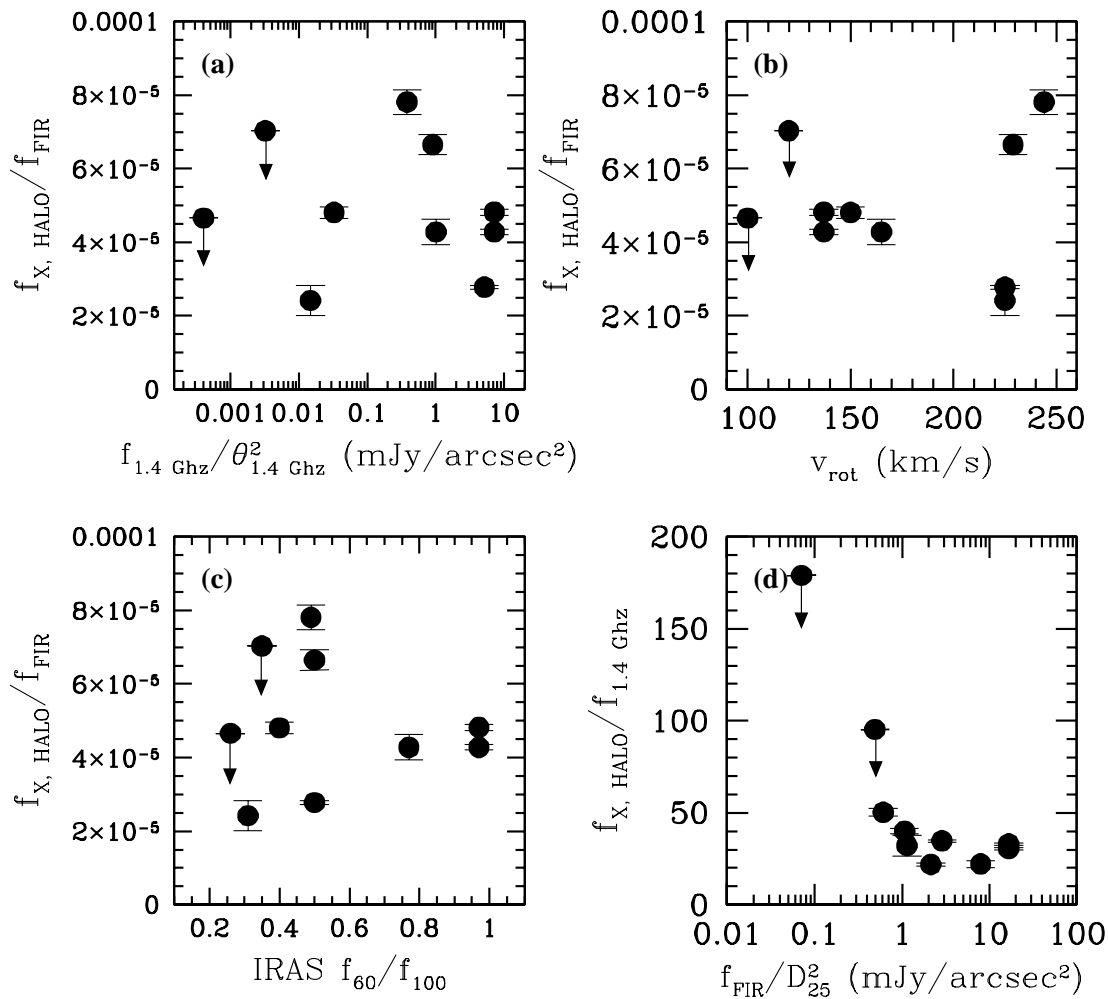}
\caption{Panels a through d plot the ratio of the absorption-corrected 
  0.3--2.0 keV halo-region
  X-ray flux to the host galaxies IRAS FIR flux (a, b and c) or
  1.4 GHz flux (d) against various proxies for the mean star formation
  rate per unit area in the host galaxy (panels a, c and d) and 
  galaxy mass (panel b).
  Error bars show the  $1\sigma$ statistical errors due to uncertainties in
  the background-and-point source subtracted count rates alone -- systematic
  uncertainties in $f_{\rm X, HALO}$ are probably a factor 2 or so.
  For NGC 6503 and NGC 4244
  the $3\sigma$ upper limits are shown.
  For NGC 253 the extrapolated total halo-region flux is used.
  These results imply that the ratio of the halo X-ray luminosity
	to the star formation rate of the host galaxy is independent
	of the concentration or intensity of the star formation,
	or the mass of the host galaxy.
  }
\label{fig:halo_fluxratio}
\end{figure*}

In Fig.~\ref{fig:halo_fluxratio} we plot the ratio of the 
absorption-corrected X-ray flux from the halo to the host galaxy's
FIR or radio flux, against estimators of the star-formation rate per
unit disk area from radio and FIR observations, and the galaxy
circular velocity. From panels a and b of this figure,
it is apparent that all galaxies with detected extra-planar
X-ray emission have a similar $f_{X, HALO}/f_{\rm FIR}$ ratio, of
approximately $4\times10^{-5}$ ($\pm{0.2}$ dex). The mechanical
energy injection rate from supernovae and stellar winds in
the starburst is $L_{\rm SN}\sim 0.01 \, \epsilon \times L_{\rm FIR}$,
where $\epsilon$ is the fraction of mechanical energy thermalized
\citep{ham90}. This implies that the halos of these starburst galaxies,
along with the normal spiral galaxy 
NGC 891, radiate an approximately fixed fraction 
$\sim 0.004/\epsilon$ of the
mechanical energy supplied by stellar feedback within the disk, 
in the 0.3--2.0 keV X-ray band. For any realistic value of $\epsilon$,
\ie $0.1 \le \epsilon \le 1$, we re-derive the well-known result that
superwinds radiate a negligible fraction of their 
energy budget in soft X-rays (we discuss this 
further in \S~\ref{sec:discussion:blow_out:therm_effic}). 

Perhaps the most interesting implication of
Fig.~\ref{fig:halo_fluxratio} is that the X-ray luminosity per unit
star formation rate is independent of the star formation intensity
(star formation rate per unit disk area), over many orders of
magnitude variation in star formation rate per unit area.
In other words, the efficiency of converting supernova mechanical
power into X-ray emission appears to be 
independent of the original supernova
rate per unit area or volume.

It is also interesting that the normal spiral NGC 891 is so similar
to the starbursts in this regard. Fig.~\ref{fig:halo_fluxratio} also
makes it clear that the non-detections of diffuse emission in the
halos of NGC 6503 and NGC 4244 are not particularly constraining -- given
their low FIR and radio brightness we would expect any diffuse X-ray
emission generated by star-formation activity to be as faint or fainter
than our existing $3\sigma$ limits.

\begin{figure*}[!ht]
\epsscale{0.9}
\plotone{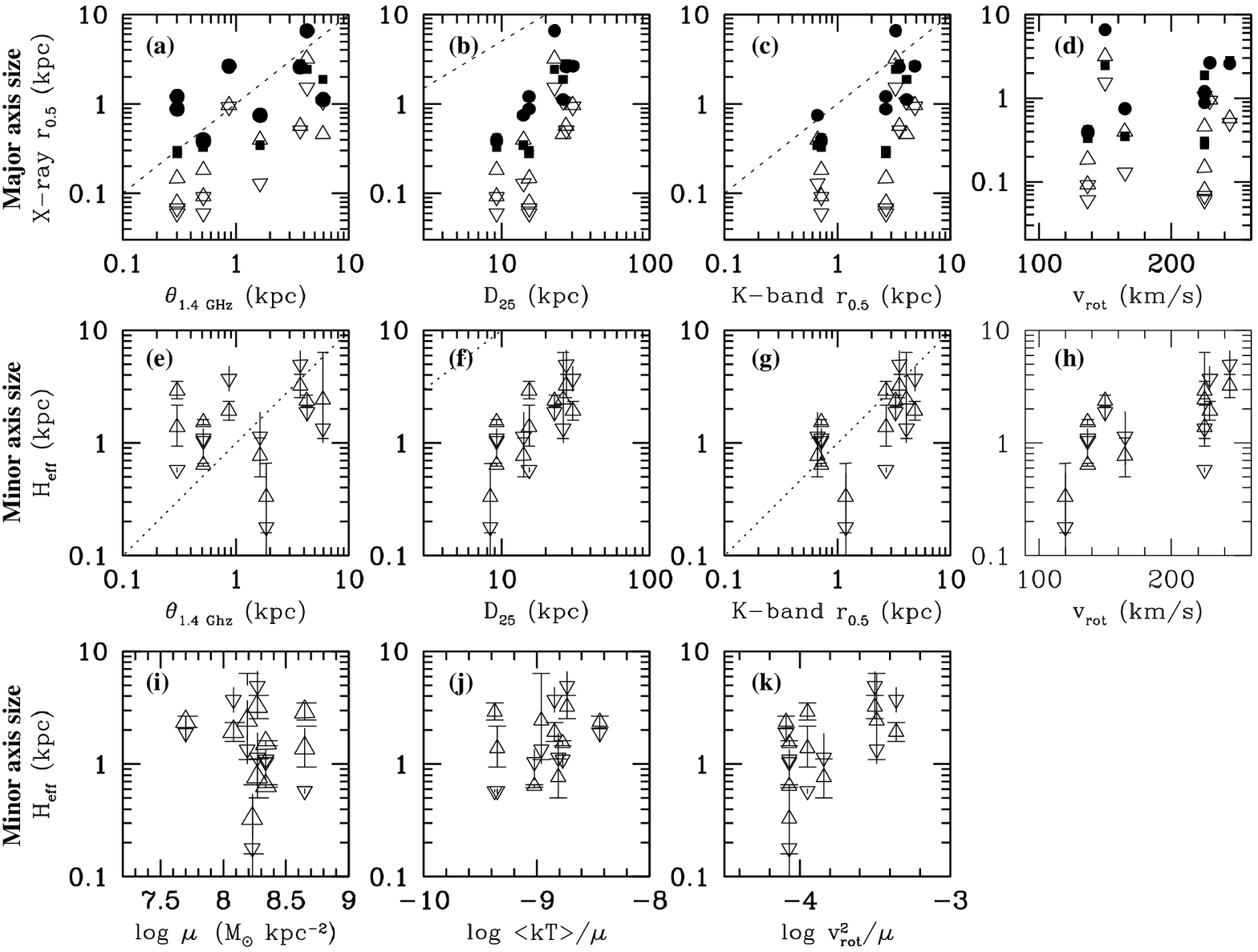}
\caption{A comparison of the characteristic physical dimensions of the
diffuse X-ray emission to various proxies for the size of the
star-burst region, the optical size of the host galaxy, and galaxy mass
(the half-light radius of the 1.4 GHz radio emission
$\theta_{1.4 GHz}$, the inclination corrected $D_{25}$ value from the 
RC3 catalog, the K-band half-light radius from the 2MASS Large
Galaxy Atlas \citep{2mass_largegals}, and 
the circular rotation velocity $v_{\rm rot}$).
In panels a to d these values are plotted against the 1.0--2.0 keV energy band
major-axis diffuse x-ray half-light and 75\%-enclosed light
radii $r_{0.5}$ (open symbols) and $r_{0.75}$ (filled symbols), while in
panels e through h they are plotted against the
minor axis diffuse X-ray exponential scale height $H_{\rm eff}$ 
in the 0.3--1.0 keV energy band.
Data measured along the positive $r$ or $z$-direction are plotted as 
open triangles ($r_{0.75}$ values are shown as filled circles), 
negative $r$ or $z$ as inverted triangles ($r_{0.75}$ values 
are shown as filled squares). The dotted lines in panels a -- c
and e -- g
show the trend an exact one-to-one correspondence between the variables
would produce. See discussion in \S~\ref{sec:discussion:mass_or_sf:sizes}.
In panels i through k we plot the vertical scale height against the
logarithm of mean mass surface density $\mu$
(in $\Msol$ kpc$^{-2}$), the ratio of the mean X-ray temperature
to the mass density (units of keV kpc$^{2}$ $\Msol^{-1}$), 
and the ratio of the square of the circular velocity
to the mass density (km$^{2}$ s$^{-2}$ kpc$^{2}$ $\Msol^{-1}$).}
\label{fig:sizes_sizes1}
\end{figure*}

\subsection{The size of the X-ray emitting regions}
\label{sec:discussion:mass_or_sf:sizes}

The major-axis X-ray half-light radii $r_{0.5}$ (tabulated in 
Table~5 of Paper I) provide a convenient measure
of the radial size of the diffuse X-ray emission in and near the plane of each 
galaxy (as a large fraction of the diffuse X-ray emission comes from 
$z \la 2$ kpc of the plane). 
To reduce the influence of absorption on the measured
sizes we plot $r_{0.5}$ derived from the medium energy band (1.0--2.0 keV)
against various simple measures of the radial extent of 
strong star-formation ($\theta_{\rm 1.4 GHz}$), 
the extent of the old stellar population ($D_{25}$ and the 
K-band $r_{0.5}$), and the galactic mass (circular velocity $v_{\rm rot}$)
in Fig.~\ref{fig:sizes_sizes1}a -- d.
The radial extent measure $r_{0.5}$ correlates weakly with
$\theta_{\rm 1.4 GHz}$, $D_{25}$ and K-band $r_{0.5}$, and 
not at all with $v_{\rm rot}$. The 75\%-enclosed light radius $r_{0.75}$
shows a somewhat better correlation with $D_{25}$ and the K-band $r_{0.5}$.
Quantitatively, 
the diffuse X-ray radial extent $r_{0.5}$ is
closest to the size of the non-thermal radio emission $\theta_{\rm 1.4 GHz}$,
\emph{as  expected if the diffuse X-ray emission is generated by mechanical
feedback from massive stars}. The X-ray emission is
more centrally concentrated than the radio emission, although this may
due to the low spatial resolution of the NVSS data.

In contrast, the vertical extent of the diffuse X-ray emission in
the halo appears independent of $\theta_{\rm 1.4 GHz}$.
We have only shown
the effective halo exponential scale height $H_{\rm eff}$ 
in Fig.~\ref{fig:sizes_sizes1}e -- h, but the other tracers of minor axis
extent, such
as $z_{0.95}$ or $z_{1e-9}$, produce very similar looking plots.
Instead of correlating with local measures of where the 
energy is injected (such as $\theta_{\rm 1.4 GHz}$), 
the distribution of X-ray-emitting gas 
at $z \ga 2$ kpc depends on the total size of the host galaxy, 
given the approximately linear relationship
between $H_{\rm eff}$ and  $D_{25}$ and/or K-band $r_{0.5}$.
More quantitatively, Kendall's Tau statistic \citep{numerical_recipes} is only 
$\tau = 0.21$ for the correlation between $\theta_{\rm 1.4 GHz}$
and $H_{\rm eff}$ ($1.3\sigma$ from zero correlation, equivalently
probability of the correlation being spurious $p_{s} = 0.19$), whereas
the stronger correlations mentioned above have
$\tau = 0.61$ between $D_{25}$ and $H_{\rm eff}$ ($3.8\sigma, 
p_{s} = 0.00017$),
$\tau = 0.43$ between the K-band $r_{0.5}$ and $H_{\rm eff}$
($2.7\sigma, p_{s} = 0.0077$) and $\tau = 0.55$ between
$v_{\rm rot}$ and $H_{\rm eff}$  ($3.4\sigma, p_{s} = 0.00076$).

Panel j of Fig.~\ref{fig:sizes_sizes1} plots $H_{\rm eff}$ against
the ratio of the luminosity-weighted 
mean halo temperature $<kT>$ to the mean mass 
surface density in the disk $\mu = M_{\rm TF}/D_{25}^{2}$ 
(note that the vertical component of the gravitational 
force $g_{\rm z}$ is proportional to $\mu$).
The correlation not particularly strong, in fact with a $\tau = 0.46$
($2.8\sigma$ for zero correlation,  $p_{s} = 0.0045$) it is weaker
than the correlations between $H_{\rm eff}$ and $D_{25}$ or $v_{\rm rot}$.
This suggests that the X-ray temperature $<kT>$ 
and scale height $H_{\rm eff}$
are not as directly related as would be the case for simple
vertical hydrostatic equilibrium where $H \propto kT/\mu$. 

We also investigated the relationship between $H_{\rm eff}$ and the ratio of
$v_{\rm rot}^{2}/\mu$ (panel k of Fig.~\ref{fig:sizes_sizes1}). If the
vertical support of the gas now visible in X-ray emission was originally
due to semi-turbulent gas motions, then we might expect the vertical
velocity dispersion to be proportional to the rotational velocity
of the galaxy (as found for the stellar vertical velocity dispersion,
\eg \citealt{vanderkruit99}). Evidence in support of this scenario is lacking
in the current data, as the correlation is weak, $\tau = 0.31$ 
($1.9\sigma, p_{s} = 0.059$).

\clearpage
\begin{figure*}[!ht]
\epsscale{0.9}
\plotone{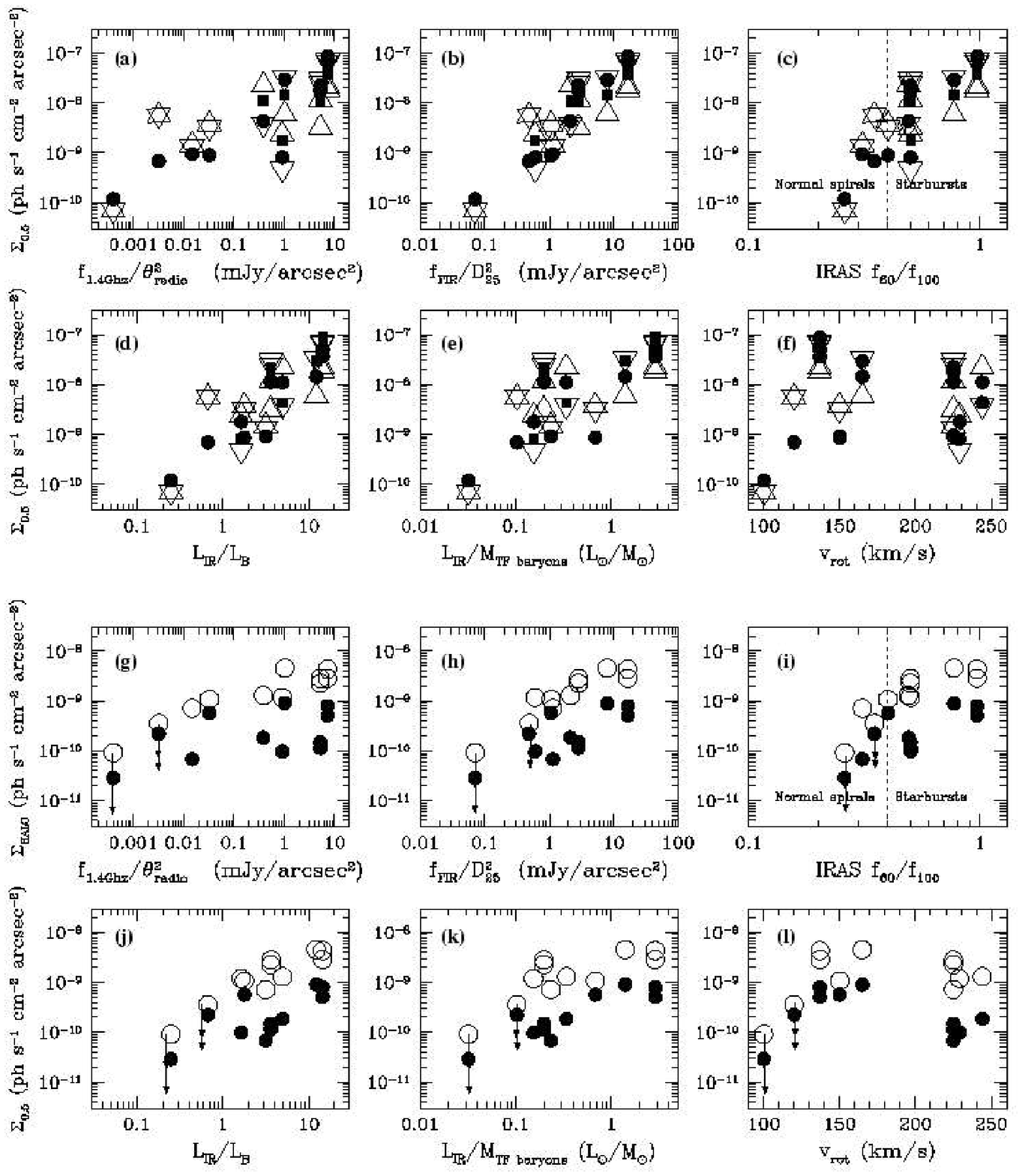}
\caption{Diffuse X-ray surface brightness plotted against various
proxies for the intensity of star formation (star formation rate per unit 
disk area from $f_{\rm 1.4 GHz}/\theta^{2}_{\rm 1.4 GHz}$, 
$f_{\rm FIR}/D^{2}_{25}$ or $\log f_{60}/f_{100}$), total star-formation
rate per unit galaxy mass ($L_{\rm IR}/L_{B}$ or somewhat more 
accurately $L_{\rm IR}/M_{\rm TF}$), and galaxy mass 
(as $M_{\rm TF} \propto v_{\rm rot}^{3.51}$). Panels a -- f plot the
effective X-ray surface brightness $\Sigma_{0.5}$ within the 
half-light height $z_{0.5}$, which is largely a measure of the brightness
near or within the disk. Open symbols correspond to the 0.3--1.0 keV
energy band, filled symbols to the 1.0--2.0 keV energy band. Triangles
and squares represent data measured along the positive-$z$ direction,
inverted triangles and circles data measured along the negative-$z$ 
direction. Panels g -- l plot the mean halo region surface brightness
 in the 0.3--1.0 keV band (open circles) 
and the 1.0--2.0 keV band (filled circles). The six-sided stars are cases
where values from positive and negative $z$ overlap.
See discussion in \S~\ref{sec:discussion:mass_or_sf:surfb}.}
\label{fig:surfb_sfintens}
\end{figure*}

\subsection{The brightness of the diffuse X-ray emission, compared to 
star-formation intensities and galaxy masses}
\label{sec:discussion:mass_or_sf:surfb}

The presence of correlations between the halo X-ray luminosity and 
total star-formation rate, and the vertical extent of the X-ray-emitting
gas to the total size of the host galaxy, should automatically lead to
correlations between the diffuse X-ray surface brightness and
estimates of the star formation rate per unit area (star formation rate
intensity, SFRI) in the host galaxy.

This is indeed what we find, as shown in Fig.~\ref{fig:surfb_sfintens}.
In panels a through e we plot a variety of proxies for the
SFRI, and in panel f the circular velocity,
against the mean diffuse X-ray surface brightness within the half-light height
$\Sigma_{0.5}$.
Panels g through l repeat this, this time using the halo region
surface brightness $\Sigma_{\rm HALO}$ 
obtained within the fixed physical-size apertures used in the
halo-region spectral fitting.

Although the half light heights are typically only $\sim 1$ kpc or so,
 $\Sigma_{0.5}$ 
is still a measure of the brightness
of gas associated with the minor axis outflow in the starburst galaxies
of this sample. We have demonstrated
in this and previous papers that the brightest diffuse X-ray emission 
is directly associated with optically-emitting gas for which 
unambiguous kinematic evidence of minor-axis outflow exists 
\citep[\eg, in][]{ham90,mckeith95,shopbell98}. In the normal
spirals, NGC 891, NGC 6503 and NGC 4244, there is no existing
kinematic evidence for outflow, and hence one might expect
$\Sigma_{0.5}$ to be more of a measure of
 the brightness of the hot gas {\em within} 
the disk (as opposed to lying in projection within 2 kpc of the mid-plane),
and also the mean brightness of unresolved point sources.

Note that
the correlation between $\Sigma_{0.5}$, or $\Sigma_{\rm HALO}$, 
and the FIR SFRI $f_{FIR}/D_{25}^{2}$ (Fig.~\ref{fig:surfb_sfintens}b and h, 
where the angular size $D^{2}_{25}$ is that of the old stellar 
population, and hence independent of the angular size of the main
star forming region in the host galaxy) 
are better than the correlations between
 $\Sigma_{0.5}$ or $\Sigma_{\rm HALO}$ and the NVSS 1.4 GHz SFRI  
$f_{\rm 1.4 GHz}/\theta_{\rm 1.4 GHz}^{2}$ 
(Fig.~\ref{fig:surfb_sfintens}a and g, where $\theta_{\rm 1.4 GHz}$ 
is a direct, although low spatial resolution, measure 
of the size of regions in which SNe
are occurring). This is due to the better correlation between the
vertical extent of the extra-planar emission with the total size 
of the host galaxy, than with the radial size of the main 
star-formation regions (as discussed in 
\S~\ref{sec:discussion:mass_or_sf:sizes}).

The ordinate values in panels d, e, j and k of Fig.~\ref{fig:surfb_sfintens} 
are, to first order, star formation rate per unit galactic mass.
These panels, along with  panels f and l (which plot $\Sigma_{0.5}$ and
$\Sigma_{\rm HALO}$ against the 
circular velocity $v_{\rm rot}$), show that the brightness of the
diffuse X-ray emission does not correlate with galactic mass, and hence
is genuinely
\emph{caused} by star formation. 
It is therefore not the case that the correlation
in panels a through c, and g through h, is spurious and arises
 simply because more massive galaxies, which can
hold more hot gas in their potential wells, also have more
star-formation in total.

Again, it is intriguing that there is no obvious transition
in X-ray surface brightness between the normal and starbursting
spirals -- the normal spirals follow the trends one could extrapolate
on the basis of the starburst galaxies alone. The sample of non-starburst
disk galaxies we have available to compare against the starbursts 
is very small, but no good reasons exist as yet to believe that these
galaxies are peculiar.

In summary, these 
results validate the basic assumption that the presence of extra-planar
diffuse X-ray emission is driven by star-formation processes in the
underlying disk, but that the large-scale structure of the host galaxy
also plays a strong role in determining the properties and
distribution of the resulting extra-planar
emission.

\begin{figure*}[!ht]
\epsscale{0.9}
\plotone{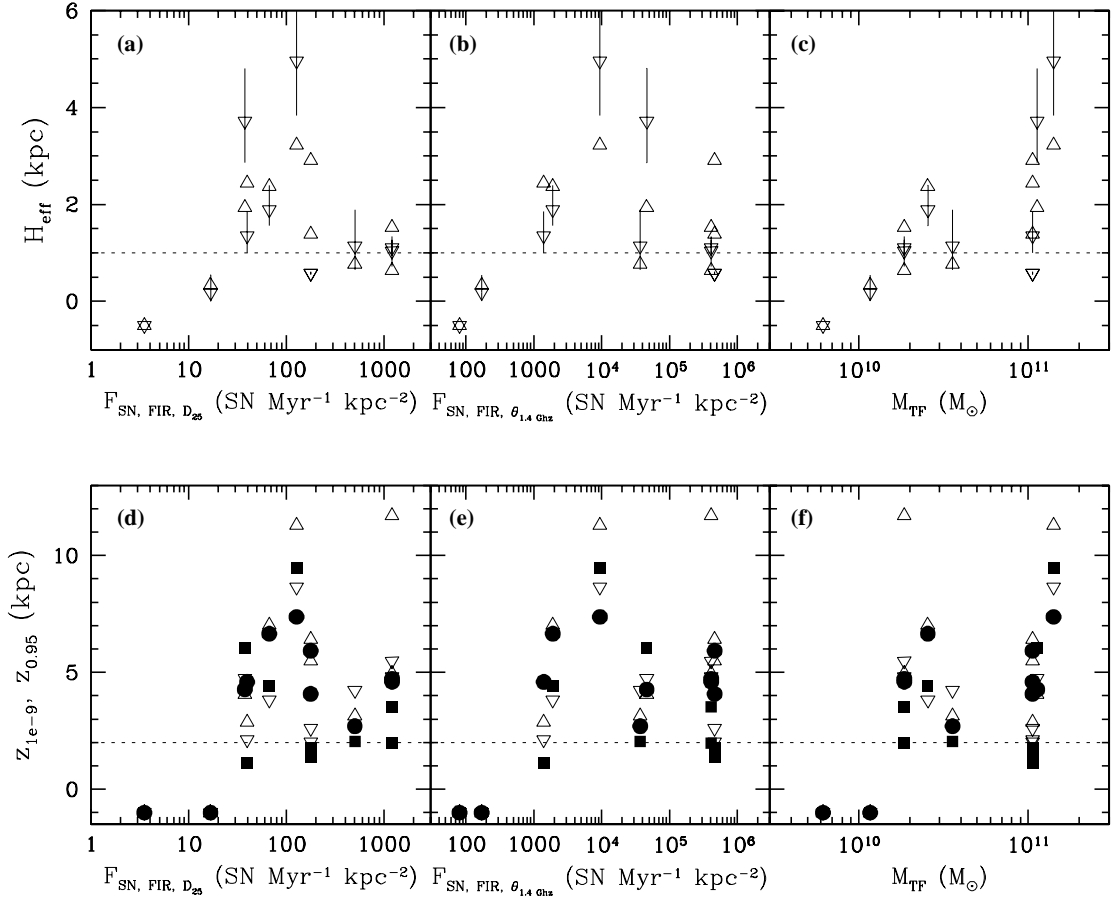}
\caption{The vertical extent of the diffuse X-ray emission
	as a function of the mean SN rate per unit area $F_{\rm SN}$ or
	circular velocity $v_{\rm rot}$ of each galaxy. 
	See \S~\ref{sec:discussion:blow_out:comparison} for 
	a description of how $F_{\rm SN, FIR, D_{25}}$ 
	and $F_{\rm SN, FIR, \theta_{\rm 1.4 GHz}}$ were calculated.
	Panels a -- c plot the fitted surface brightness scale height
	for $z \ge 2$ kpc, data from positive $z$ shown as open triangles
	and negative $z$ as inverted triangles. The horizontal
	dashed line corresponds to a gas density scale height of $2$ kpc.
	Panels d -- g plot the isophotal minor axis size $z_{1e-9}$
	(positive $z$ data points shown as open triangles, 
	negative $z$ as inverted triangles) 
	and 95\% enclosed minor axis light height $z_{0.95}$ 
	(positive $z$ data points shown as filled circles, 
	negative $z$ as filled squares).
	Where no
	measurement can be made a negative value is shown.}
\label{fig:observed_fsn}
\end{figure*}

In Fig.~\ref{fig:observed_fsn} we show the observed vertical extent
of the diffuse X-ray emission against estimates of the SFRI
(recast in terms of supernova rate per unit area $F_{\rm SN}$) and
total galactic mass $M_{\rm TF}$ for
the galaxies of our sample. We estimate the supernova rate based on the
IRAS FIR flux. Dividing this SN rate by 
the inclination corrected $D^{i}_{25}$ value
gives $F_{\rm SN, FIR, D_{25}}$, the mean SN rate per unit area 
of the entire disk. If instead we use the approximate diameter of the
non-thermal radio emission $\theta_{\rm 1.4}$ we obtain 
 $F_{\rm SN, FIR, \theta_{\rm 1.4 GHz}}$ 
(See Tables~\ref{tab:galaxies} and \ref{tab:galactic_sf} for the 
origin of these values). It is clear that
\emph{there is no correlation between the vertical extent of the
diffuse X-ray emission and the SN rate per unit area in the host galaxy},
except that galaxies with lower SN rate per unit area than NGC 891 have
no detected extra-planar X-ray emission. This is similar to other studies
of extra-planar emission at radio and optical wavelengths.
\citet*{dahlem95} found no correlation between the
vertical scale height of non-thermal radio emission and the
star formation rate per unit area in a sample of edge-on disk galaxies
\citep[see also][]{dahlem01}. In a \halpha~imaging study of
edge-on disk galaxies \citet{miller03} also find no
correlation between scale heights and the star formation rate per unit disk
area.

Fig.~\ref{fig:sizes_sizes1}, 
in combination with Fig.~\ref{fig:observed_fsn}, 
implies that the density scale height
of the X-ray emitting material depends
on the absolute size of the host galaxy, and not on the spatial
distribution or intensity of star formation within it.
Note that this is not a statement regarding the absolute physical
extent of superwinds, which almost certainly is much greater
than the region probed by current observations (and which
will physically depend on the total
amount of energy deposited by SNe and the density of the IGM).


We were initially motivated to investigate the quantitative
relationship between
the extra-planar X-ray properties and the SFRI indicators  
$L_{\rm FIR}/D_{25}^{2}$ and $f_{\rm 1.4 GHz}/\theta^{2}_{\rm 1.4 GHz}$ 
by optical studies of the eDIG
\citep[see][ among others]{dettmar92,rand96,hoopes99,rossa03}. 
These had demonstrated the presence of extra-planar
optical emission correlated with $L_{\rm FIR}/D_{25}^{2}$, in
the sense that galaxies with high values of $L_{\rm FIR}/D_{25}^{2}$ 
were more likely to show evidence of eDIG. Combined with the
studies of \citet{lehnert95,lehnert96,dahlem01} that demonstrate
that IR-warm galaxies with $f_{60}/f_{100}> 0.4$ are extremely
likely to have extra-planar emission, kinematic evidence of outflow
and/or non-thermal radio halos, we expected a correlation
between  high star formation rate per unit area and the luminosity and
extent of diffuse X-ray emission in disk galaxy halos.

Nevertheless, in the X-ray band it appears that the correlation we
have found between the surface brightness of the diffuse X-ray
emission and star formation rate per unit area 
is the combination of two independent effects:
\begin{enumerate}
\item A relatively uniform efficiency of converting mechanical energy
  to X-ray emission (\S~\ref{sec:discussion:mass_or_sf:rates}), \ie
  $f_{\rm X} \propto f_{\rm FIR}$, 
  the exact physics of which is not fully understood.
\item The vertical extent of the X-ray emission is directly
  related to the size of the host galaxy, $H_{\rm eff} 
  \propto D_{25}$ or the K-band half-light radius
  (\S~\ref{sec:discussion:mass_or_sf:sizes}), 
  possibly for the reasons discussed below. 
\end{enumerate}

This does not mean that the SFRI is not an important parameter, in particular
when considering the conditions necessary to drive gas into the
halo of a galaxy (as we shall discuss below), but that 
other parameters play a significant role in shaping the emission we see.

In terms of the interpretation of the extra-planar
X-ray and \halpha~emission presented in Paper I
(and summarized in \S~\ref{sec:paper1_summary}), item 2 above
can be explained if disk galaxies have gas throughout their halos,
with vertical density scale heights approximately similar to
the optical diameter of the host galaxy. 
Depending on the exact model assumed the mass of gas
in the halo of a Milky Way-like galaxy \emph{prior} to any starburst
is a few times $10^{7}$ to a few times $10^{8} \Msol$
(see models 4 and 5 in \citealt{strickland02}).

\subsection{Limits placed on cosmological accretion models for the
origin of hot gas in the halos of normal galaxies}

Without deep X-ray observations of a larger sample of edge-on normal
spiral galaxies it is difficult to critically test cosmological 
accretion models of the type presented
by \citet{toft02}, and \citet{sommerlarsen02}. Excluding the
starbursts from consideration for the time being being, we would
expect halo 0.3--2.0 keV X-ray luminosities of order 
$L_{\rm X} = 4.60$, 0.20 and  $0.08\times10^{38} \ergps$ 
for NGC 891, NGC 6503 
and NGC 4244 respectively, based on the
\citeauthor{toft02} model.

These luminosities are based on converting the bolometric luminosity as
a function of rotation velocity quoted in \citet{toft02} into {\it Chandra}
energy band soft X-ray luminosities.
Here we have assumed the X-ray emission has a fixed temperature of
0.16 keV (the virial temperature for the NGC 891 halo 
$kT_{\rm vir} = \mu m_{\rm H} v_{\rm rot}^{2}$.
The observed luminosity-weighted mean temperature obtained for 
NGC 891 is $<kT> = 0.23$ keV), a metal
abundance $Z=0.3\Zsol$ (the conversion between the $L_{\rm BOL}$ used
by \citeauthor{toft02} 
and 0.3--2.0 keV energy band $L_{\rm X}$ is largely independent
of $Z$), and is
described by an exponential surface brightness profile with
a fixed surface brightness scale height of 2 kpc (as \citealt{toft02}
do not provide any measure of how the characteristic size of the
X-ray emission varies with galaxy rotation velocity). For such a
distribution a fraction $e^{-1}$ of the total luminosity arises in
gas at $|z| \ge 2$ kpc. \citet{toft02}'s predicted bolometric
luminosity (0.012-12.4 keV energy band) of $L_{\rm BOL} \sim 10^{40} \,
(v_{\rm rot}/230 \kmps)^{5} \ergps$ (scatter $\pm{0.5}$ dex)
thus converts to a total 0.3--2.0 keV X-ray luminosity (\ie integrated
over all heights $z$) of 
$L_{\rm X} \sim 1.4\times10^{39} \, (v_{\rm rot}/230 \kmps)^{5} 
\ergps$ (for simplicity we ignore the fact 
that the halo temperature should vary 
with rotation velocity as approximately $kT \propto v_{\rm rot}^{2}$,
as hence even lower X-ray fluxes would be expected in the
0.3--2.0 keV energy band for NGC 6503 and NGC 4244).

The predicted accreted halo luminosities 
given in the first paragraph of this
section should be compared 
the to the observed values of $L_{\rm X, HALO} = 1.2 \times 10^{39} \ergps$ 
for N891 and $3\sigma$ upper limits of $L_{\rm X, HALO} < 1.6 \times 10^{38}
\ergps$ and $< 2.5 \times 10^{37} \ergps$ 
for NGC 6503 and NGC 4244 respectively (see Table~9 of Paper I). 
This comparison indicates that the predicted accretion-model
X-ray luminosity for a galaxy like NGC 891 
is close to that value observed, and that
the NGC 4244 and NGC 6503 observations are not deep enough to strongly
constrain the accretion model. It is worth noting that it is far from
clear what fraction of the gas accreted from the IGM by disk galaxies 
is heated to the halo virial temperature 
(see for example \citealt{katz03} and references therein for a theoretical
perspective).

Although, based on luminosity alone,
some fraction of the X-ray emission in the halo of NGC 891 might be 
due to gas accreted from the IGM, we find the reasoning presented in
\S~\ref{sec:discussion:mass_or_sf:rates}, 
\ref{sec:discussion:mass_or_sf:sizes} and 
\ref{sec:discussion:mass_or_sf:surfb} 
more compelling as evidence that the extra-planar 
X-ray emission in NGC 891 is associated with, and dominated by, feedback
from star-formation.
This apparent commonality of normal spirals with starbursts,
based on the properties of their extra-planar emission,
might be an artifact of the small number statistics. 
Deep X-ray observations of other edge-on normal spirals,
spanning a broad range of star-formation rates, are clearly necessary
before firmer conclusions can be drawn. Nevertheless, a robust finding
of a smooth  transition in the halo gas properties from normal
to starburst galaxies would have major implications for understanding
the large-scale structure of the ISM and stellar feedback processes 
in galaxies.

\clearpage

\section{Comparison the theoretical models of mechanical energy feedback
from massive stars}
\label{sec:discussion:theory}

\subsection{Disk blow-out energy requirements}
\label{sec:discussion:blow_out}
In a seminal paper, \citet{maclow88} considered, both analytically and
numerically, the amount
of mechanical energy that supernovae and stellar winds need to
supply for a superbubble to blow-out of the
stratified atmosphere of a galactic disk. Blow out occurs if the dense
shell of swept-up gas marking the outer boundary of a 
superbubble ceases to decelerate
and begins to accelerate down the density gradient. 
At this point Rayleigh-Taylor instabilities set in and rapidly fragment the
superbubble shell, leading to the venting of the hot high pressure gas
from the interior of the bubble out into the surrounding medium. As this
venting of hot gas (the gas blowing out) occurs when the outer shell fragments
and breaks up the term ``breakout'' is often used synonymously with blow out. 
This theory, with minor
alterations, has been widely used since that time, in particular
with respect to assessing the
the prospects for metal ejection from galaxies 
\citep*[\eg][]{koo92,martin96,maclow96,tt96,silich98,maclow99,ferrara_tolstoy00,fps00,silich01}. Note that blow out of a superbubble from the 
disk ISM of a galaxy does not automatically imply that gas escapes into
the IGM --- more mechanical work must be done by the vented hot gasses
to move any gaseous halo medium. Simulations of superwinds show that these
hot gasses blow new bubbles in the halo, sweeping the ambient halo gas up
into new shells \citep[\eg][]{tomisaka93,suchkov94,ss2000}. 
We shall address this 
in \S~\ref{sec:discussion:ejection}.

\citeauthor{maclow88} demonstrate that whether or not a bubble blows/breaks
out of a given atmosphere depends only on
 the mechanical energy input rate, \ie mechanical luminosity, driving
the bubble, and the pressure and scale height of the ambient ISM the bubble
is expanding into --- for convenience we re-derive this result below.
For such an important theoretical construct, there has been surprisingly
little in the way of observational testing and calibration.
\citet{ham90} demonstrated
that the energy injection rates in starburst galaxies exceeded
this critical blow-out luminosity (hereafter $L_{\rm CRIT}$), 
but is is not clear from that work (and other comparisons to
starbursting galaxies \eg \citealt{martin96}) what the lowest
mechanical energy injection rate is that leads to blow-out in real galaxies.

In this section we shall use the observationally determined presence
of hot gas in the halos of our sample galaxies as empirical
evidence for blow-out from the disk, 
and compare the estimated rates of SN energy
return in disks of these galaxies to theoretical values of $L_{\rm CRIT}$.
At present the observational results do not strongly constrain the theory.
When larger samples of edge-on spirals have had deep X-ray observations made
of them the promise of this method should be realized.

\subsubsection{The critical mechanical luminosity for disk blow-out}
\label{sec:discussion:blow_out:lcrit}
The theoretical critical mechanical luminosity required for blow-out is
based on the following concepts. If, in a stratified disk with a 
vertical gas density distribution $\rho_{z} = \rho_{0}\,
\exp(-z/H)$, an expanding superbubble or SN remnant has an expansion
velocity $v_{\rm bub}$ (the shell expansion velocity)
greater than the sound speed $c_{\rm s} = (\gamma
 P / \rho)^{1/2}$
in the ambient ISM by the
time it reaches a vertical height of order the scale height $H$ of
the disk, then the dense shell will begin to accelerate. At this
time the bubble or remnant shell will fragment due to Rayleigh-Taylor
instabilities. The hot, high pressure, plasma interior to the
shell will vent out through the fragmented shell into the surrounding
medium, \ie the halo for blow out from the disk of a galaxy.
The dense fragments of the superbubble shell will be carried out
along with the hot gas, in fact the ram pressure of the hot gas streaming
past them will accelerate the shell fragments to higher velocities
than the original shell expansion velocity $v_{\rm bub}$
\citep[see \eg][]{chevclegg,ham90,heckman00}.

If instead $v_{\rm bub} \la c_{s}$, the bubble's expansion will continue to
decelerate and eventually coast to a stop without blowing out.
Radiative cooling will eventually lead to depressurization of the bubble
interior, followed by the collapse of the bubble, in addition to
damage caused by galactic differential rotation and random ISM motions.

A more formal analysis identifies $v_{\rm bub} > c_{\rm s}$ at $z=3H$
as the critical requirement for blow-out in a single-phase 
exponential atmosphere \citep[see][]{ferrara_tolstoy00}, numerically 
verified by simulations
showing blowout occurring once the bubble reaches a height of 
$\sim 3H$ \citep*{maclow88,maclow89}

For a bubble driven by a constant mechanical luminosity $L_{\rm W}$ into
a medium of uniform and constant 
ambient density $\rho_{0}$ \citep*{castor75,weaver77},
the radius $r_{\rm bub}$ and shell expansion 
velocity $v_{\rm bub}$ as a function of time $t$ are
\begin{equation}
r_{\rm bub} = \alpha \, L^{1/5}_{\rm W} \, \rho^{-1/5}_{0} \, t^{3/5},
\label{equ:rbub}
\end{equation}
and
\begin{equation}
v_{\rm bub} = 0.6\, \alpha \, L^{1/5}_{\rm W} \, \rho^{-1/5}_{0} \, t^{-2/5},
\label{equ:vbub}
\end{equation}
where $\alpha = (125/154\pi)^{1/5}$. Rearranging these equations gives us
the expansion velocity
at the point that the bubble radius is 3H:
\begin{equation}
v_{\rm 3H} = 0.6 \times 3^{-2/3} \, {\cal F}_{v} \, \alpha^{5/3} \,  L^{1/3}_{\rm W} \, 
	\rho^{-1/3}_{0} H^{-2/3},
\label{equ:v3h}
\end{equation}
where ${\cal F}_{v}$ is a numerically-determined correction factor
to account for the vertical decrease in density leading to
bubble expansion velocities greater than in the constant density 
case. Using a code that numerically integrates the 
appropriate equations of motion
\citep{ss99}, we find ${\cal F}_{v} = 
2.5$\footnote{\citet{ferrara_tolstoy00} present
a purely analytical derivation of $v_{\rm 3H}$ 
including the effect of the density
decline, but we find this method over-predicts $v_{\rm 3H}$ by a factor 3.7
over our numerical result (even when errors in their equations
24 and 25 are corrected for). We speculate that this discrepancy is
due to their neglect of the inertia of the cold shell in their calculation
of the shock velocity.}.

The resulting minimum mechanical luminosity required to blow out of this
atmosphere, $L_{\rm CRIT}$, is
\begin{equation}
L_{\rm CRIT} = (154\pi/3) \, \gamma^{3/2} \, {\cal F}^{-3}_{v} 
	\, \rho_{0}^{-1/2} \,
	P^{3/2}_{0} \,H^{2},
\label{equ:lcrit}
\end{equation}
where $P_{0}$ is the thermal 
pressure in the ambient ISM. Note that this luminosity is
a factor $\sim 3$ lower than the critical 
luminosity derived by \citet{maclow88},
\ie blow-out from the disk is slightly easier.
Converting to units more appropriate for
the disk of a normal spiral, 
\begin{equation}
\frac{L_{\rm CRIT}}{10^{38} \ergps} = 3.43 \, n_{0}^{-1/2} 
	\left(\frac{P_{0}}{10^{4} k}\right)^{3/2} 
	\left(\frac{H}{\rm 1 kpc}\right)^{2},
\label{equ:lcrit_scaled}
\end{equation}
where $n_{0} = \rho_{0}/\mu m_{\rm H}$ and $\mu m_{\rm H} = 
10^{-24} {\rm g} \pcc$.
More energy is needed to blow out of
a starburst region, where measured ISM pressures reach $P/k \sim 10^{7}$ 
K cm$^{-3}$ \citep{mccarthy87,ham90,lehnert96}.

\subsubsection{Disk blow out due to individual stars clusters, or through 
cooperation between multiple clusters?}

With respect to blow out from the disk of
normal star forming galaxies is it commonly assumed that
each superbubble is powered by a single star cluster. In contrast,
it has long been recognized that superwinds from galaxies with
nuclear starbursts are powered by
the collective energy return from the massive stars within the
entire starburst region, typically a few hundred parsecs in
size. This is a well established observational result that can be demonstrated
in various ways, \eg the superwind pressure profiles in \citet{ham90} and
\citet{lehnert96} demonstrate that mass and energy are being injected
relatively uniformly over a region equal in size to the starburst
region. Another example is that 
the edges of best-resolved superwind outflows, in galaxies such as
M82 and NGC 253, matching up well to the edges of the 
nuclear starburst regions \citep{ss2000,strickland00}. 
This is by no means surprising,
as star clusters tend themselves to cluster spatially just as
massive stars tend to form within clusters, with
starbursts representing the extreme of the phenomenon (one can describe the
M82 starburst as $\sim 100$ R136's within a region only 2--3 times larger
than 30 Doradus, see \citealt{oconnell95}). Furthermore, \citet{meurer95}
demonstrate that only a fraction $\sim 20$\% of the massive stars within 
nearby starburst regions are associated with very 
massive clusters, there being a
substantial population of massive stars in 
starburst regions not associated with obvious
clusters \citep[see also][]{tremonti01,chandar03}. 

The question arises whether superbubbles can be cooperatively
driven by massive stars in multiple stellar clusters 
in situations less extreme than those in nuclear starbursts, \ie
in disk-wide starbursts or even in normal star forming galaxies?
This process need not be as efficient or as common as in the nuclear
starbursts, but it might occur often enough to be significant in
terms of the disk/halo interstellar interaction. 
There are a number of reasons why it is worthwhile to 
consider such a scenario. Firstly, individual
clusters that are massive enough to drive a superbubble capable
of blowing out of the disk of a normal spiral galaxy on their
own are relatively rare, as we show in
\S~\ref{sec:discussion:blow_out:noninteract}.
If only one or two such clusters exist in a galaxy at any given time
it would be difficult to sustain a galaxy-wide fountain-driven 
hot halo (assuming that these are not extremely peculiar spiral galaxies).
Secondly multi-dimensional simulations of the ISM with energy return
by SNe and stellar clusters show that bubbles do merge, and
once merged can potentially become powerful enough to escape the disk
\citep{avillez00}.
A third issue, related to the first, is that the normal
spiral galaxy NGC 891 has a very similar
ratio of halo-region X-ray luminosity to FIR luminosity (\ie total star
formation rate) to the starbursts with superwinds, implying a similar
total efficiency in converting the initial supernova mechanical energy into 
soft X-ray emission. 
If cooperative action does not occur in normal spirals and only
the most massive cluster drive bubbles that blow out of the disk, then a
smaller fraction of the massive stars forming in NGC 891 would be
responsible for the hot gas its the halo. To achieve a 
$f_{\rm X, HALO}/f_{\rm FIR}$ ratio similar to that in the
starbursts would then require fine tuning the physical 
processes in the halo of NGC 891 to be more efficient at radiating
soft X-rays, despite the other similarities to the extra-planar
X-ray emission from the starbursts. 

A final reason to consider blow out due to collective superbubble driving is
that it naturally leads to an analytical 
formulation in which there is a critical
star formation rate (or supernova rate) per unit area necessary for blow out. 
From the stand point of observational studies of 
galaxies beyond the local group 
it is much easier to estimate a mean star formation per unit area than to
measure the properties of individual stellar clusters. Indeed,
individual star clusters are likely be heavily obscured in the edge-on
galaxies where the presence of extra-planar hot gas is empirical evidence
for blow out from the disk. Furthermore estimates of the star formation rates
in these galaxies may only be available 
using low spatial resolution FIR observations.
As we shall discuss later, this formulation remains useful
when  considering blow out from the halo even if cooperative action does
not apply to superbubble blow out from the disk.

\subsubsubsection{Energy return from non-interacting star clusters}
\label{sec:discussion:blow_out:noninteract}
Let us now consider what the mass is of an OB association that contains enough
massive stars enough to power a superbubble capable of blowing out 
of a disk of a \emph{normal spiral 
galaxy}, assuming that only a single cluster can power
a specific superbubble. The critical
energy given in Eqn.~\ref{equ:lcrit_scaled} can be recast in terms of
the number of massive stars formed in the cluster, and from
that in terms of star cluster mass. Assuming all stars in a cluster
form instantaneously, then the period over which energy injection occurs
corresponds to the lifetime of the least massive star to undergo
core collapse, i.e. stars with initial masses $M_{i} \sim 8 \Msol$,
which is $\sim 30$ -- $40$ Myr. The energy injection rate over this period
is approximately constant, see \eg \citet{lh95}.
The time needed for a bubble with $L_{\rm W} = L_{\rm CRIT}$
to expand to the point it breaks out of the disk 
is similar to this time period, 
validating the assumption of continous energy injection in the formulae
used to derive $L_{\rm CRIT}$.
We assume that each SNe
returns $\epsilon \times 10^{51} \erg$ of mechanical 
energy to the ISM, where the thermalization 
efficiency $\epsilon$ is the fraction 
of the initial kinetic energy per supernova
that can be used to drive a bubble, \ie the energy
fraction not radiated
away prior to each young remnant overlapping with other SN remnants.
We further assume a Salpeter IMF with upper and lower mass limits
of 100 and $1 \Msol$ respectively, which gives one massive star per
$\sim 53 \Msol$ of stars formed. Where we to assume a more realistic
IMF extending down to the hydrogen burning limit 
\citep*[\eg that of][]{kroupa93} we would get less SNe per mass of stars
formed, \ie the cluster masses required for disk blow out would be larger
and hence blow out rarer.

The Milky Way has a gaseous scale height of $\sim 0.4$ -- 
$0.5$ kpc in the neutral
ISM \citep{dickey90,sembach94}, and between 0.5 -- 1.5 kpc for the
free electrons tracing the warm ionized component 
\citep*{reynolds89,nordgren92}. The mean number density and thermal pressure
in normal star-forming spiral like the Milky Way or NGC 891 are
$n_{0}\sim1\pcc$ and $P/k \sim 10^{4}$ $\K \pcc$ 
\citep{ferriere01,wolfire03}. 
Assuming $H=0.5$ kpc, $n_{0}= 1 \pcc$ and $P/k = 10^{4}$ $\K \pcc$ 
then $L_{\rm CRIT} = 8.6\times10^{37} \ergps$ in a typical region of a 
normal spiral galaxy. This correspond to the energy return from
$109 \epsilon^{-1}$ SNe, or a cluster mass of 
$M_{\rm CL} \sim 5.8\times10^{3} \epsilon^{-1} \Msol$.
A canonical  value of the supernova 
thermalization efficiency under \emph{normal} interstellar conditions
 is $\epsilon = 0.1 $ (\citealt{thornton98}, 
note that $\epsilon$ is expected to be higher
in regions of higher star formation rate per unit volume such as
star clusters \eg \citealt{larson74,canto00}).
These results are consistent with those of \citet{fps00},
who also considered blow out from the disk assuming
that superbubbles can only be blown by a single star cluster.
For galaxies with star formation rates equivalent to the normal spiral
galaxies we are discussing they derived a critical number of SNe 
for blow out similar to, or larger than, the number we derive above.

Star clusters as massive as $M_{\rm CL}$ are not common in normal,
non-starbursting, galaxies. By way of
comparison consider the
masses of the \emph{most spectacular and massive star clusters in 
the Local Group}.
The Arches cluster in the Galactic center,
although containing $\sim 5$\% of all the Wolf-Rayet stars in 
the Galaxy, has a mass $<7\times10^{4} \Msol$ \citep{figer02}.
The R136 cluster in the 30 Doradus nebula of the LMC has an
estimated initial mass of $\ga 1.7 \times 10^{4} \Msol$ 
\citep{malamuth94}, or $\sim 3 \times 10^{4} \Msol$, 
upper limit $1.5 \times 10^{5} \Msol$ \citep{brandl96}.
The entire Orion Nebula Cluster,
has a total gravitating mass of $4.5\times10^{3} \Msol$ 
\citep{hillenbrand98}. 
From this it is clear that the most massive,
most populous, star clusters are capable of driving superbubbles powerful
enough to blow out of normal spiral galaxy disks on their own,
unless $\epsilon$ is of order unity.

How common are such clusters, and what fraction of massive stars
form in these massive clusters? The number of massive stars per 
star cluster appears to follow a power law such that the number of
of clusters with $N$ massive stars is proportional to $N^{-2}$
\citep[see][and references therein]{oey03}. We will treat this as
a continuous distribution with limits $N_{\rm MIN}=1$ and $N_{\rm MAX}$
for computational convenience, as the results from the exact
solution \citep[see][]{oey03} 
are essentially identical in the regime we consider. The fraction
of star clusters containing $N$ or more 
massive stars is thus $f_{>N}= (N^{-1}-N_{\rm MAX}^{-1})/(1-N_{\rm MAX}^{-1})$.
If we generously assume that even normal galaxies can potentially form star 
clusters with masses up to $10^{6} \Msol$, then 
$N_{\rm MAX} \sim 1.9 \times10^{4}$. Assuming a lower $N_{\rm MAX}$ 
will reduce $f_{>N}$. For
cluster thermalization efficiencies of 0.1, 0.3 and 1.0 
the required number of SN for disk blow out are
1090, 327 and 109 respectively, with corresponding cluster masses
of $M_{\rm CL} = 6 \times 10^{4}, 1.8\times10^{4}$ and $6\times10^{3}
\Msol$. For these three efficiencies the fraction of star clusters
powerful enough to blow out of the disk of a normal spiral
galaxy are $f_{>N} = 8.7\times10^{-4}, 3.0\times 10^{-3}$ and 
$9.1\times10^{-3}$ respectively. In other words it is rare to find a
star cluster that is capable of driving, \emph{on its own}, a superbubble that
can blow out of the disk of a normal spiral galaxy.

\subsubsubsection{Collective energy return over galactic scales}
\label{sec:discussion:blow_out:interact}
Motivated by the preceding arguments we explore the requirements
for blow out from the disk of normal spirals
in a scenario at the opposite extreme from 
isolated, non-interacting star cluster,
model commonly adopted. We shall assume that separate star
clusters can cooperatively blow a single bubble, just as 
starburst regions drive a single superbubble and then superwind.
This gives rise to a \emph{critical star formation rate 
(or supernova rate) per unit area necessary for blow out}, which is
a number that is easier to empirically constrain 
in edge-on disk galaxies (\eg see Table~\ref{tab:galactic_sf}) 
than the masses of specific clusters.

\citet{maclow88}
find that at blowout the radius of the superbubble within the
disk is $\sim H$. We
therefore assume that some significant fraction $\Upsilon$ 
of all SNe due to massive star formation
within an area $\pi H^{2}$ of the disk
can contribute energetically to creating
a blowout, \ie.
\begin{equation}
L_{\rm W} = \pi H^{2} \, F_{\rm SN} \, \epsilon \, \Upsilon \, E_{\rm SN},
\label{equ:fsn}
\end{equation}
where $F_{\rm SN}$ is the SN rate per unit disk area, and 
$E_{\rm SN} = 10^{51} \erg$ is the kinetic energy return per SN event.

Note that hydrodynamical simulations  demonstrate that 
the expansion of a wind-blown bubble or superbubble is
relatively insensitive to the spatial distribution of the 
energy injection within it \citep[\eg][]{ss98,ss2000}, 
for the reasons discussed by 
\citet{maclow88}. We can thus continue to use the standard expansion
laws despite having a more randomly distributed source of mechanical power.

Physically $\epsilon$ is the thermalization efficiency within 
individual clusters where radiative losses are small \citep[\eg][]{canto00},
while $\Upsilon$ represents the effective efficiency of combining the
energy output from the multiple clusters within the $\pi H^{2}$ area
of disk that can possibly contribute to a bubble just about to blow out.
Theoretically, both $\epsilon$ and $\Upsilon$ are functions of the
 star formation
rate, as supernovae alter the porosity of the ISM, and
hence alter radiative energy losses and the
effectiveness of collective action \citep{larson74}. Relatively little
theoretical work has been done with the aim of estimating $\epsilon$
for realistic cluster conditions, and this theoretical work has not
been tested against observation in any meaningful way. Although
simulations of the
ISM that include interacting superbubbles do exist \citep[\eg][]{avillez00},
these tend not to discuss the energetic efficiency with which separate
bubbles merge.  One situation somewhat similar
to that we envisage can be found in a numerical simulation of a primordial
galaxy by \citet*{mori02}. They simulated bubbles blown 
by 90 star clusters spread over a region $\sim 1$ kpc in radius
and measured the fraction of the total mechanical energy injected
by SNe escaping into the IGM to be $\sim 30$ -- $50$\%, depending on
the initial conditions used. 
Their simulation effectively assumed $\epsilon =1$, and their galaxy is
not exactly comparable to a present day spiral disk (\eg it lacks
a halo), but these results
imply $\Upsilon > 0.3$ -- $0.5$ (in our model $\epsilon\Upsilon$ is 
the fraction of available SN mechanical power released within the disk
that is available to drive the final bubble that blows out, but less 
than this fraction of energy will reach the halo 
due to radiative energy losses that would occur 
even in the absence of superbubble merging and collective action).
One would expected $\Upsilon$ to be of this order in normal
disk galaxies. Physically the primary radiative energy losses in two
merging bubbles would come from the colliding shells, 
as these are dense and hence will
be efficient radiators if shocked in the collision. However, only $\sim20$\%
of the total mechanical energy originally fed into a superbubble is thought to
be contained in the shell, the majority of the energy ($\sim 45\%$) 
residing in the hot interior and $\sim35$\% being lost during shell formation 
\citep{maclow88}. 
The hot gas in merging bubbles is unlikely to suffer significant 
additional radiative losses, given its low density.
For starbursts, where the evidence for effective collective action
is best, $\Upsilon$ may be of order unity, as the higher SN rate per
unit volume should lead to higher ISM porosity and hence
efficient thermalization \citep[\eg][]{larson74}. In other words
the net efficiency $\epsilon\Upsilon$ throughout the entire starburst
region might be comparable to the thermalization efficiency immediately
within a dense star cluster \citep[\eg][]{canto00}. By way of comparison
the SN rate per unit volume in the center of M82 is approximately
two orders of magnitude higher than the SN rate per unit volume in
the simulations of \citet{mori02} that we discuss above.

Combining Eqn.~\ref{equ:fsn} with Eqn.~\ref{equ:lcrit}, we obtain the critical
SN per unit disk area for blow out from the disk, $F_{\rm SN, CRIT}$, to be
\begin{equation}
F_{\rm SN, CRIT} = (154/3) \, \gamma^{3/2} \, {\cal F}^{-3}_{v} 
	\, \rho_{0}^{-1/2} \,
	P^{3/2}_{0} \, (\epsilon \, \Upsilon \, E_{\rm SN})^{-1}.
\label{equ:fcrit}
\end{equation}
\emph{Note that this is no longer dependent on the scale height. Large gaseous
scale heights do not inhibit blow out as the bubble can draw mechanical
energy from a correspondingly larger region of disk.} 
Scaling to more convenient units,  
\begin{equation}
F_{\rm SN, CRIT} ({\rm Myr}^{-1} {\rm kpc}^{-2}) = 3.45 \times  n_{0}^{-1/2} 
	\left(\frac{P_{0}}{10^{4} k}\right)^{3/2} \, 
	\left( \frac{1}{\epsilon \Upsilon}\right),
\label{equ:fcrit_scale}
\end{equation}
or alternatively
\begin{equation}
F_{\rm SN, CRIT} ({\rm Myr}^{-1} {\rm kpc}^{-2}) = 3.45 \times  
	\left(\frac{P_{0}}{10^{4} k}\right) 
	\left( \frac{1}{\epsilon \Upsilon}\right) 
	\left( \frac{T_{0}}{10^{4}} \right)^{1/2}.
\label{equ:fcrit_scale:alt}
\end{equation}

For the purposes of comparing observation with theory
 $F_{\rm SN}$ should ideally be calculated on a location-by-location
basis within a given galaxy, based on the local supernova rate
${\cal R}_{\rm SN}$, or star formation rate, summed
within an area $\pi \, H^{2}$, where $H$ is the
appropriate gas density scale height for the region we are
interested in assessing blow out from
(be it the thin molecular disk, the more extended Reynolds layer,
or the halo itself)
--- \ie some knowledge of the scale height is still required.

\subsubsection{Constraints on feedback efficiencies from the
observed luminosities}
\label{sec:discussion:blow_out:therm_effic}

Both the standard bubble blow out theory presented 
in \S~\ref{sec:discussion:blow_out:lcrit}
and the slightly modified version above depend strongly on
the efficiency of mechanical energy feedback through
terms such as $\epsilon$ and $\Upsilon$. Theoretical
work suggests these efficiencies depend on 
the local star formation rate through the porosity of
the ISM \citep{larson74},
but the exact values are hard to quantify and there is
little agreement between different groups 
(\eg contrast \citet{strickland_vulcano} with \citet*{recchi_vulcano}).

Adopting a different approach, we can 
crudely constrain the net mechanical energy feedback efficiency
by relating the rate of mechanical energy injection into the halo
to the observed X-ray luminosities.
From the relationship between the FIR luminosity and SN rate given
in \citet{ham90}, $R_{\rm SN} = 0.2 L_{\rm FIR}/10^{11}$ \Lsol,
we obtain that the rate of mechanical energy injection into the
halo is
\begin{equation}
L_{\rm W, HALO} \sim 0.0167 \, \epsilon\Upsilon \, L_{\rm FIR},
\label{equ:lmech}
\end{equation}
where we ignore energy losses due to superbubble
shell formation and any superbubble
merging that occurs (see \S~\ref{sec:discussion:blow_out:interact}).
Including these radiative losses might reduce $L_{\rm W,HALO}$ by
a factor $\sim 2$, and lead to a similar increase in the estimated
efficiencies, if analytical theory and numerical models of
winds are right \citep[see \eg][]{ss2000}.

Direct observation of the volume-filling high-pressure gas that
drive superwinds would be the best method of measuring the
supernova thermalization efficiency. In the absence of this
we are forced to estimate $\epsilon$ from 
a theoretically-determined relationship between the
X-ray luminosity and the mechanical power actually driving the wind.
The fraction $A_{\rm XRAD} = L_{\rm X, HALO}/L_{\rm W, HALO}$
depends on the physical process responsible for the
soft X-ray emission, \eg non-radiative wind shocks, or possibly
conductive interfaces between hot and cold gas. 
For X-ray emission from conductive interfaces \citep{weaver77,chu90}, which is
the mechanism that has been most extensively explored to date,
$A_{\rm XRAD}$ depends only weakly on the ambient density 
and the value of the thermal conductivity ($L_{\rm X}$ scales almost
exactly in proportion to $L_{\rm W}$, and increases weakly
for higher ambient density or thermal conductivity).
We calculated $A_{\rm XRAD}$ for conditions appropriate for
superwinds using the code described in \citet{ss99}, which numerically
solves the \citet{weaver77} model and uses the \citet{mekal} hot plasma
code to calculate the X-ray emission. For $L_{\rm W, HALO}$ in the
range $10^{40}$ to $10^{42} \ergps$, gas densities in the range
$10^{-3}$ to $1 \pcc$, ages in the range $5$ to $30$ Myr and
values of the thermal conductivity in units of the Spitzer
conductivity 0.1 to 10, we find a mean $A_{\rm XRAD} = 0.013$ (with a
scatter $\pm{0.6}$ dex). In our hydrodynamical simulations of
superwinds \citep{ss2000}, which do not include thermal conduction, we
find a similar fraction of the mechanical energy emerging in the
soft X-ray band (again with a significant scatter, depending on the
exact initial conditions assumed).

For the time being we shall assume that $A_{\rm XRAD}\sim 0.01$
is a reasonable estimate of the fraction of  
the wind mechanical power that is radiated in the soft X-ray band.
Combining this with Equation~\ref{equ:lmech} we obtain
\begin{equation}
L_{\rm X, HALO} \sim 1.67\times10^{-4} \, (A_{\rm XRAD}/0.01) \, 
\epsilon\Upsilon \, L_{\rm FIR}.
\label{equ:lxray}
\end{equation}
The typical $f_{\rm X, HALO}/f_{\rm FIR}$ ratio we observe
is $\sim 4\times 10^{-5}$, with the range being $2.2$ -- $8.0\times 10^{-5}$, 
which this implies $\epsilon\Upsilon \sim 0.24$ with a range of
0.14 -- 0.48.
The normal spiral galaxy NGC 891 has one of 
the lowest $f_{\rm X, HALO}/f_{\rm FIR}$ ratios in
the sample, with $f_{\rm X, HALO}/f_{\rm FIR} = 2.2\times10^{-5}$, and thus
a lower than average value for $\epsilon\Upsilon \sim 0.14$
(NGC 891 is the lowest if we exclude the 
somewhat doubtful detection of halo emission from NGC 4945, 
although it is only a factor $\sim 2$ lower than the 
mean $f_{\rm X, HALO}/f_{\rm FIR}$ ratio).
a lower than average value for $\epsilon\Upsilon \sim 0.14$.

\subsubsection{Comparison to observations}
\label{sec:discussion:blow_out:comparison}

If, as seems most probable, the origin of the diffuse extra-planar
X-ray emission around disk galaxies is due to SN feedback, 
then empirically assessing whether significant blow out from the disk
has occurred is simple. It effectively  reduces to 
identifying which galaxies have diffuse
X-ray emission at $z \ga 2$ kpc (\ie several disk scale heights). 
With a sample of
edge-on galaxies, some of which have extra-planar
hot gas and other which do not, and with reasonable
estimates of the star formation or SN rate per unit area 
for each galaxy we can 
then empirically determine (a) whether there is a
critical SN rate per unit area ($F_{\rm SN,CRIT}$)
above which blow out from the disk
occurs, and if so (b) the value of this critical SN rate per unit area.

In principle we could compare the empirically derived value of
$F_{\rm SN,CRIT}$ with the theoretical value 
(\eg Equ.~\ref{equ:fcrit_scale}), and hence test whether
the theory is correct or not. 
In practice it is currently impossible to use this method to
meaningfully test superbubble blow out theory due to a mixture of
theoretical and observational uncertainties.

The primary theoretical uncertainties are in the 
mechanical energy thermalization efficiency for individual star clusters 
$\epsilon$ and efficiency of cooperation between multiple clusters
$\Upsilon$. The largest uncertainties in these parameters
exist the for star forming conditions in.normal disk galaxies, \ie
galaxies that probably span the critical threshold in star formation rate
per unit area for disk blow blow out.

Observational difficulties include uncertainties in estimating the
star formation or supernova rates, and more significantly the 
uncertainties in the effective area 
and hence the mean SN rate per unit area.

Currently it is only practical to estimate the total supernova
rate ${\cal R}_{\rm SN}$ of each of these galaxies (or other edge-on
disk galaxies that will be observed with {\it Chandra} or {\it XMM-Newton}). 
This prevents us from calculating
$F_{\rm SN}$ in the location-by-location manner described in 
\S~\ref{sec:discussion:blow_out:interact}\footnote{We note that \halpha~imaging
could be used to evaluate $F_{\rm SN}$ on a location-by-location
basis within the same galaxy, although this method has
typically been used on to calculate mean galactic SFRI values
and blow-out thresholds \citep[\eg][]{rossa03}.
\halpha~emission in the disk is typically much more
radially extended than the regions responsible 
for driving the outflow
(the exception being M82, where star-formation is now confined solely to the
nuclear region), \ie star formation in some regions of the disk
is occurring at rates per unit area below the critical value.
This is graphically illustrated in Fig.~\ref{fig:rgb_circ} and 
\ref{fig:rgb_halo}. Of course, \halpha~measurements suffer
the weakness that they systematically underestimate
the true SFRI in the regions with the highest star formation rates 
due to the associated higher levels of extinction.}. Nor do we have exact
measures of the scale height of the disk ISM, and $F_{\rm SN}$ depends
sensitively on the assumed scale height.
Considering
the issue of disk blow out, the ISM scale
height is likely to be of order
a few hundred parsecs to a kiloparsec or two.
As this is similar to the magnitude of $\theta_{\rm 1.4 GHz}$ 
(see Table~\ref{tab:galactic_sf}), one could argue that
it is reasonable to
use ${\cal R}_{\rm SN}/(4 \theta_{\rm 1.4 GHz}^{2})$ 
as a plausible $F_{\rm SN}$ to use
when assessing disk blow 
out (this value is denoted as $F_{\rm SN, FIR, \theta_{\rm 1.4 GHz}}$
in Table~\ref{tab:galactic_sf} and in Fig.~\ref{fig:observed_fsn}).
In contrast the $D_{25}$ diameter (typically $\sim 10$ -- $20$ kpc) 
is commonly used when estimated
mean galactic star formation rates per unit area in optically-based
studies of extra-planar warm ionized gas 
\citep[\eg][]{rand96,hoopes99,collins00,rossa03,miller03}. This
leads to estimated mean SN rates per unit area orders of
magnitude lower (this value is 
denoted as $F_{\rm SN, FIR, D_{25}}$
in Table~\ref{tab:galactic_sf} and in Fig.~\ref{fig:observed_fsn}).

From  Fig.~\ref{fig:observed_fsn} it appears that all galaxies with
$F_{\rm SN, FIR, \theta_{\rm 1.4 GHz}} \ga 2000$ SN
Myr$^{-1}$ kpc$^{-2}$ have diffuse extra-planar X-ray emission,
indicative of disk blow out. Let us assume that the divide between disk blow
out and superbubble confinement occurs at or near this value of $F_{\rm SN}$.
As the galaxy with the lowest $F_{\rm SN}$ value that also has
\emph{detected} extra-planar emission is the normal spiral galaxy NGC 891,
let us assume that $n_{0} \sim 1 \pcc$, $P/k \sim 10^{4} \K \pcc$
and $\epsilon\Upsilon \sim 0.14$. Thus equation~\ref{equ:fcrit_scale} would
then predict disk blow out for $F_{\rm SN} \ga 25$ SN Myr$^{-1}$ kpc$^{-2}$,
a factor $\sim 80$ lower than the observationally-estimated value, apparently
a strong discrepancy between blow out theory and observations.
However, if we use $F_{\rm SN, FIR, D_{25}}$ instead, then galaxies
with $F_{\rm SN, FIR, D_{25}} \ga 40$ SN
Myr$^{-1}$ kpc$^{-2}$ appear to have disk blow out,
apparently in good agreement with blow out theory. 

In summary, although it is relatively easy to empirically determine
that a galaxy has blown hot gas out of the disk into the halo, 
a quantitative comparison to theory is much more difficult.
With the present uncertainties it is not clear whether the energy requirements
for superbubble blow out that the standard theory predicts (with or without
allowing for collective energy return from multiple clusters)
are accurate, even
to within an order of magnitude.
More theoretical and observational study is required before
we can meaningfully test superbubble blow out theory:
\begin{enumerate}
\item X-ray observations must be deep enough to rule out significant
  amount of hot gas in the halos of low SF rate galaxies, so that we
  can be certain that blow out from the disk has not occurred 
	(see \S~\ref{sec:discussion:mass_or_sf:rates}).
\item Observational estimates of star formation or supernova rates
  per unit area should ideally use spatially resolved estimates of the
  star formation rate, and more importantly, observed disk gaseous scale 
	heights (see \S~\ref{sec:discussion:blow_out:interact}).
\item We need more theoretical study of the energetic efficiencies of
  mechanical feedback in clusters and in the realistic multi-phase
  ISM, in order to understand how $\epsilon$ and $\Upsilon$ depend
  on the star formation intensity and local ISM properties
	(see \S~\ref{sec:discussion:blow_out:therm_effic}).
\end{enumerate}

\subsection{Ejection of metal-enriched gas into the IGM by superwinds}
\label{sec:discussion:ejection}

Much of the preceding discussion as revolved around the blow out of
superbubbles from the disk of normal star forming galaxies. We now
wish to briefly turn to the issue of ejection of hot, metal-enriched,
gas into the IGM from starburst-driven superwinds.

\subsubsection{The plausibility of halo blow-out}
\label{sec:discussion:ejection:halo_blowout}

Material blowing out from the disk 
does not automatically escape into the IGM, as it must work against
the obstacle provided by gas in the halo before being able to reach the
IGM, as mentioned in \S~\ref{sec:introduction}. 

Based on the observed X-ray surface brightness profiles of the
extra-planar emission of the galaxies within this sample, it is
appropriate to represent the original halo gas distribution as
an exponential with density scale height $H_{\rm g} \sim 4$ -- $8$ kpc, 
\ie twice the observed surface brightness scale height. Note that
optical studies of the extra-planar emission from warm ionized gas in
normal and some starburst galaxies also indicate exponential gas 
distributions with very similar scale heights 
\citep{hoopes99,collins00,miller03}.

For such a stratified halo gas distribution the standard superbubble blow out 
energy requirements can be applied (this time with
respect to halo blow out) to any bubble or wind that manages to blow out
of the disk. Simulations of superwinds clearly show that the halo
medium is swept-up into a new shell surrounding the hot gas vented from
the disk \citep{suchkov94,ss2000}. 
We will assume that superbubble blow out theory is
correct for the remainder of this section.
Note that this directly implies that
the critical region of space around a galaxy
that can prevent escape into the IGM is within $\sim3H_{\rm g}$
($\sim12$ -- 24 kpc) 
of the mid-plane, a region that can be probed by existing observational
methods.
We need not concern ourselves about the density and pressure of gas at
larger, $\sim 100$ kpc, distances from the host galaxy\footnote{
This is true unless
the galaxy happens to be in a high density environment such as a
cluster or compact galaxy group, where the pressure and density of the
inter-cluster or inter-group medium may be significant, in which case
the arguments of  \citet{silich98,babul,silich01} apply.}.

We can now apply the arguments of \S~\ref{sec:discussion:blow_out} 
to halo blow out. The proportionality
between the halo region X-ray flux and
 the IRAS FIR flux (and hence the \emph{total} 
galactic SF rate) strongly suggests that the halo can draw on
mechanical energy from a significant fraction of the entire star-forming
disk. If the SN rate per unit area in the disk is sufficient to
blow out from the disk, then the collective blow-out model can be applied
to the hot gas now in the halo. 
The appropriate scale height to use is that of the
halo gas density distribution $H_{\rm g}$, either directly
measured from the X-ray data or estimated using empirical results of
\S~\ref{sec:discussion:mass_or_sf:sizes}. There we showed
that $H_{\rm g} \propto D_{25}$
(or the K-band half light radius $r_{0.5}$). 
Thus the appropriate mean supernova rate per unit area 
$F_{\rm SN}$ to calculate
for assessing halo blow out is proportional to 
${\cal R}_{\rm SN}/D_{25}^{2} \equiv F_{\rm SN, FIR, D_{25}}$.

The only significant differences from the case of assessing disk blow out
are then as follows:
\begin{enumerate}
\item The appropriate density and
pressure are that of the halo, evaluated at the galactic mid-plane
(\eg see \citealt{shull_halo}).
For typical galactic environments now and at high redshift, the density
and pressure of any bound, non-transient, 
halo medium will be less than those in disk.
\item The additional energy losses during disk blow must be
accounted for. In the starbursts, a good case can be made that 
radiative losses are
negligible, based both on the observational lack of significant X-ray,
\ion{O}{6} and optical emission \citep{ham90,heckman01,hoopes03}.
In the high resolution hydrodynamical simulation of \citet{ss2000}
$\ga 50$\% of the mechanical energy injected within the disk
is transported into the halo.
\item When considering
blow out from the halo it is appropriate to use a SN rate averaged 
over the characteristic size of the host galaxy, not just the main star
forming region. For comparison to Eqn.~\ref{equ:fcrit} a good estimate
of $F_{\rm SN}$ would be 
${\cal R}_{\rm SN}/(\pi H^{2}_{\rm g}) \approx 
{\cal R}_{\rm SN}/(4\pi r^{2}_{\rm 0.5, K-band})$ 
or ${\cal R}_{\rm SN}/(0.04 \pi D^{2}_{25})$, given the correlations
shown in Fig.~\ref{fig:sizes_sizes1}.
\end{enumerate}

Making full use of this to assess blow out from the halos of galaxies 
requires that we know the critical SN rate per unit area for disk blow out.
Nevertheless, with an observationally-calibrated theory of this sort
it will be possible to assess whether gas can blow out into
the IGM. What this analysis does indicate is that extended gaseous halos 
around normal star-forming galaxies
(excluding galaxies in dense gaseous environments such as clusters)
do not automatically prevent outflows from reaching the IGM. 
The total power requirements for blowing out of a gaseous halo
will more somewhat more
exacting than for simply blowing gas out
of the disk, but this is not surprising, nor a significant 
problem given the SN rates in nuclear starbursts. 
Nevertheless the most important point we wish readers to
extract from this discussion is that the 
critical region of space around a galaxy
that can prevent escape into the IGM is most probably 
within $\sim3H_{\rm g}$ ($\sim12$ -- 24 kpc) 
of the mid-plane, which is a region well probed by existing observational
methods.

\subsubsection{Requirements for escape into the IGM}

Unambiguous observational proof of ejection of material into the IGM
by outflows
requires kinematic evidence of outward motion at $v \ga v_{\rm esc}$
at heights $z \ga 3 H_{\rm g}$ above the mid-plane of the host 
galaxy (where $v_{\rm esc}$ is the escape 
velocity at that location, and not the
escape velocity from the center of 
the halo, and $H_{\rm g}$ is the density scale 
height of the gaseous halo)\footnote{Note that $v$, and preferably the
summed thermal and kinetic energy, 
must be assessed on a phase-by-phase
basis. The cool shell velocity
\citep[as used by, for example][]{ferrara_tolstoy00,fps00}, or the temperature
of the X-ray emitting plasma \citep[\eg][]{wang95,martin_tescape}, are not
particularly meaningful tools for assessing whether gas will be retained.
The energy per particle in the hot phases is much greater than
that of the cool shell material, and it is likely that 
the kinetic energy of the X-ray-emitting
hot gas can exceed its thermal energy several times \citep{ss2000}.}
\footnote{We wish to emphasize that gas motions at velocities less than
the local escape velocity do not necessarily mean that the gas will
be retained by the host galaxy. The gaseous phases in
superwinds for which kinematics are available are entrained into
the wind, either by Kelvin-Helmholtz instabilities
along the wall of the outflow, or as fragments of the original
superbubble shell that fragmented at disk blow out.
\emph{Their motions are not ballistic}, as they are
embedded in the higher velocity fluid of merged supernova ejecta.
Following blow out from the disk their kinematics are no 
longer determined by the standard \citet{weaver77}
wind-blown bubble model.}. 
This is observationally challenging, especially given that the
most interesting, \ie metal-enriched, gas phases are very tenuous,
and very hot and highly ionized.
Such observations, even of the denser phases, are unlikely to be accomplished
for many years. Nevertheless, what the 
theoretically-and-observationally motivated argument in the
preceding section demonstrates is
that it is \emph{likely} that starburst galaxies,
even ones as massive as $M \sim 10^{11} \Msol$, and possibly some
normal spiral galaxies, do eject material into the IGM.
This is consistent with the effective yields and metal-loss
scenario presented by \citet{garnett02} and \citet{tremonti}.
Put another way, we regard escape as an eminently plausible fate for
at least some fraction of the material in superwinds, 
as currently \emph{there are no
convincing theoretically or observationally-motivated ways to absolutely
prevent escape}.

\section{Summary}
\label{sec:conclusions}

Making use of the detailed measurements of the diffuse X-ray emitting gas
in and around the sample of seven starburst 
and three normal edge-on spiral galaxies presented in Paper I, 
we have investigated how the properties of this hot gas
correlate with the size, mass, star formation rate
and star formation intensity in the host galaxies. Although not
a large sample, these galaxies do span the full range of star formation
activity found in spiral galaxies. The three normal spirals are too small
a sample  on their own
to establish any hard conclusions about the
properties of diffuse X-ray emission in normal spiral galaxies, 
but in combination with the starbursts we can make a meaningful
initial assessment of how closely the
properties of the diffuse X-ray emission in these systems resemble those of
the edge-on starburst galaxies.

We demonstrate that the extra-planar (\ie halo-region) diffuse
X-ray emission in the starburst galaxies 
is ultimately driven by star formation activity within the disk,
\ie through starburst-driven superwinds.
Accretion of gas from the IGM, 
or AGN-driven winds (see Paper I), do not
appear to be significant in this sample.

Considering both the starburst and normal spiral galaxies, 
larger galaxies show more radially and vertically extended diffuse
X-ray emission, but beyond this correlation we find no evidence that
galactic mass plays any part in determining the diffuse hot phase
in the disks and halos of these galaxies. The extent 
of the diffuse X-ray emission parallel to the plane of the host galaxy
correlates well with estimates of the extent of star formation within
the disk, further evidence that the soft thermal X-ray emission is
a result of mechanical energy feedback from massive stars. The vertical
extent of the diffuse X-ray emission correlates best with 
the absolute size of the host galaxy, as if the emitting gas fills the
halo. Exponential scale heights are similar to those found in the
extra-planar warm ionized gas from optical
for samples of star forming disk galaxies.

The luminosity of the
diffuse X-ray emission, both within the disk and halo, is determined
primarily by the rate of mechanical energy injection due to SNe
in each galaxy (which is correlated to the star-formation rate).
The surface brightness of the extra-planar 
diffuse X-ray emission correlates with
the star formation rate per unit area in the underlying disk, but
this correlation is a combination of the previously mentioned 
luminosity and size correlations.

The properties of the diffuse X-ray emission in the normal spiral
galaxies is very similar to that found in the starburst superwind galaxies.
In fact, one could successfully predict the
diffuse X-ray properties of all three normal galaxies in this sample
based on extrapolating the X-ray properties of the starburst galaxies
to the lower star formation intensities seen in the normal galaxies.
The close similarity in the halo-region diffuse 
X-ray emission in the starburst 
(with superwinds) and the Milky-Way-like normal spiral NGC 891 
(most probably having a star-formation-fed 
galactic fountain) is intriguing, and deserves more study.
Is it simply due to similar, micro-physical, emission mechanisms,
or does the similarity extend to the macroscopic hydrodynamics
at work over multi-kpc scales?
The current sample of edge-on normal spiral galaxies observed
by {\it Chandra} or {\it XMM-Newton} is small,
so it is still possible that it is purely a coincidence that 
the three normal spirals we have studied 
lie on trends extrapolated from the starbursts. With the larger
sample of observations available with next few years we should
finally be able to typify the properties of diffuse thermal X-ray emission
in normal spirals in the way we have with starburst galaxies.
 
We also consider what the observed extra-planar diffuse X-ray 
emission can tell us about the both the efficiency of massive star mechanical 
feedback on galactic scales, and the accuracy of standard theoretical
models of superbubble blow out from disk galaxies.

We find the ratio of the halo-region diffuse X-ray flux to the
host galaxy's total FIR flux is typically 
$f_{\rm X, HALO}/f_{\rm FIR} \sim 4 \times 10^{-5}$, to within
a factor two, for both the starburst galaxies and NGC 891.
This ratio is independent of the star formation rate per unit
area or the mass of the host galaxy.


It has long been clear that starburst-driven superwinds are powered
by the collective mechanical energy input from massive stars within the
entire starburst region. In contrast superbubble blow out from the
disk ISM of normal
star forming galaxies is usually
considered in the context of bubbles being powered by a single star
cluster, in which only very rare, very massive, 
star clusters are powerful enough to
drive a bubble that can blow out of a disk \citep[][]{ferrara_tolstoy00,fps00}.
We argue that there are a variety of good reasons to consider some
level of cooperative or collective action between multiple clusters
in normal star forming galaxies, \ie
when considering blow out of gas from the disk a significant fraction of
all SN energy with a disk area $\sim \pi H_{\rm g}^{2}$ can be  used
to blow a superbubble. 
We show that
a minor generalization of the \citet{maclow88}
superbubble model to allow for such collective action directly
predicts a critical star formation rate per unit area for superbubble blow
out from the disk, a formulation particularly useful for comparison to
observations of edge-on star forming galaxies.

The standard theory of superbubble blow out developed by \citeauthor{maclow88}
is widely used in astronomy, largely in relation to mass and metal
loss from galaxies, but has had little in the way of observational
testing. 
The presence of hot gas in the halo of a star forming galaxy is
direct evidence for blow out of superbubbles from the disk
of the host galaxy (in cases where the emission can be shown to be
related to SF activity in the underlying disk).
Further observational study of edge-on normal spiral
galaxies in the ways presented in this paper,
and in Paper I, will allow empirical constraints
to be made on the critical star formation rate per unit disk area
necessary to blow hot gas out of the disk into the halo.
In principle these empirical measurements could be used to test superbubble
blow out theory. 
We discuss and emphasize several observational and
theoretical issues that must be overcome before superbubble
blow out theory can be practically tested.


Finally, we apply this collective-action blow out theory to 
superwinds blowing out from the extended gaseous halos of their
host galaxies,
as our and other workers observational results
strongly suggest the gaseous halos of
star forming galaxies are well represented as exponential atmospheres.
It is important to note that 
blow out from the halo into the IGM is energetically more difficult
that blow out from the disk into the halo.
The existing uncertainties
in the accuracy of blow out theory prevent robust
quantitative assessment of whether the metal enriched gas powering
superwinds escapes into the IGM. Nevertheless, we argue that
is is still
probable that the hot, metal-enriched gas in superwinds
does escape into the IGM, even for galaxies as massive as
$M \sim 10^{10}$ -- $10^{11} \Msol$  in normal
environments (\ie not in dense galaxies groups or clusters). 
Irrespective of the quantitative
uncertainties associated with disk and halo blow out, we wish 
to emphasize that the distribution of gas in the halos of star forming 
galaxies observed in the optical and in the soft X-ray band
implies that
the crucial spatial region
around a galaxy that controls whether gas will escape into the IGM
is not the outer halo $\sim 100$ kpc from the host galaxy.
Instead it is the inner few halo scale heights, within $\sim 20$ kpc of the
galaxy plane, a region well probed by existing observational methods.

Within the next few years it will be possible to calibrate theories
of disk and halo blow out and mechanical energy feedback against observations
of local star-forming galaxies, and thus make meaningful inferences
about the enrichment of the IGM by outflows, and the influence of
massive star feedback on cosmological scales.


\acknowledgments

Over the several years this project has been in the making, we have
have been fortunate to benefit from the insightful comments and
questions of many astronomers, too numerous to mention 
individually, to whom we extend our thanks. We would also like to thank
S. Hameed for providing us with R-band and \halpha~images of NGC 1482.
We thank the anoymous referee for comments that lead to a clearer
presentation of our arguments.

DKS is supported by NASA through {\it Chandra} Postdoctoral Fellowship Award
Number PF0-10012, issued by the {\it Chandra} X-ray Observatory Center,
which is operated by the Smithsonian Astrophysical Observatory for and on
behalf of NASA under contract NAS8-39073.

This research is partially based on data from the ING Archive.
This publication makes use of data products from the Two 
Micron All Sky Survey, which is a joint project
of the University of Massachusetts and the Infrared Processing 
and Analysis Center/California Institute of
Technology, funded by the National Aeronautics and 
Space Administration and the National Science
Foundation. Furthermore, this research has made use of 
the extremely-useful NASA/IPAC Extragalactic Database (NED) which is operated 
by the Jet Propulsion Laboratory, California Institute of 
Technology, under contract with the National 
Aeronautics and Space Administration.




\begin{thebibliography}{}
\bibitem[Aguirre \etal (2001)]{aguirre01}
	Aguirre, A., Hernquist, L., Schaye, J., 
	Weinberg, D.H., Katz, N. \& Gardner, J., 2001, \apj, 560, 590
\bibitem[Armus \etal (1995)]{armus95}
	Armus, L., Heckman, T.M., Weaver, K.A. \& Lehnert, M.D.,
	1995, \apj, 445, 666 
\bibitem[de Avillez (2000)]{avillez00}
	de Avillez, M.A., 2000, \mnras, 315, 479
\bibitem[Murakami \& Babul (1999)]{babul}
	Murakami, I. \& Babul, A., \mnras, 309, 161	
\bibitem[Bell \& de Jong (2001)]{bell01}
	Bell, E.F. \& de Jong, R.S., 2001, \apj, 550, 212
\bibitem[Benson \etal (2000)]{benson00}
	Benson, A.J., Bower, R.G., Frenk, C.S. \& White, S.D.M.,
	2000, \mnras, 314, 557 
\bibitem[\protect\citeauthoryear{Bessel, Castelli \& Plez}{Bessel \etal}{1998}]{bessel98}
	Bessel, M.S., Castelli, F. \& Plez, B., 1998, \aap, 333, 231
\bibitem[Bottema \& Gerritsen (1997)]{bottema97}
	Bottema, R. \& Gerritsen, J.P.E., 1997, \mnras, 290, 585
\bibitem[Brandl \etal (1996)]{brandl96}
	Brandl, B., Sams, B.J., Bertoldi, F., Eckart, A., Genzel, R.,
        Drapatz, S., Hofmann, R., L\"owe, M. \& Quirrenbach, A., 
        1996, \apj, 466, 254 
\bibitem[Bregman (1980)]{bregman80}
	Bregman, J.N., 1980, \apj, 236, 577
\bibitem[Bregman \& Glassgold (1982)]{bregman82}
	Bregman, J.N. \& Glassgold, A.E., 1982, \apj, 263, 564
\bibitem[Bregman \& Houck (1997)]{bregman97}
	Bregman, J.N. \& Houck, J.C., 1997, \apj, 485, 159
\bibitem[Bregman \& Pildis (1994)]{bregman_and_pildis}
	Bregman, J.N. \& Pildis, R.A., 1994, \apj, 420, 570
\bibitem[Burstein \& Heiles (1982)]{burstein82}
	Burstein, D. \& Heiles, C., 1982, \aj, 87, 1165
\bibitem[\protect\citeauthoryear{Cant\'o, Raga \& Rodr\'iguez}{Cant\'o \etal}{2000}]{canto00}
	Cant\'o, J., Raga, A.C. \& Rodr\'iguez, L.F., 2000, \apj, 536, 896 
\bibitem[Carpenter (2001)]{carpenter01}
	Carpenter, J.M., 2001, \aj, 121, 2851
\bibitem[Castor, McCray \& Weaver (1975)]{castor75}
        Castor, J., McCray, R., Weaver, R., 1975, \apjl, 200, L107
\bibitem[Chandar \etal (2003)]{chandar03}
	Chandra, R., Leitherer, C., Tremonti, C. \& Calzetti, D.,
	2003, 586, 939
\bibitem[Chevalier \& Clegg (1985)]{chevclegg}
	Chevalier, R. \& Clegg, A., 1985, \nat, 317, 44
\bibitem[\protect\citeauthoryear{Chiang, Ryu \& Vishniac}{Chiang \etal}{1988}]{chiang88}
	Chiang, W.-H., Ryu, D. \& Vishniac, E.T., 1988, \pasp, 100, 1386
\bibitem[Chu \& Mac Low (1990)]{chu90}
	Chu, Y.-H. \& Mac Low, M.-M., 1990, \apj, 365, 510
\bibitem[Collins \etal (2000)]{collins00}
        Collins, J.A., Rand, R.J., Duric, N. \& Walterbos, R.A.M.,
        2000, \apj, 536, 645
\bibitem[Condon (1992)]{condon92}
	Condon, J.J., 1994, \araa, 30, 575
\bibitem[Condon \etal (1998)]{nvss}
	Condon, J.J., Cotton, W.D., Greisen, E.W., Yin, Q.F., Perley, R.A.,
	 Taylor, G.B., Broderick, J.J., 1998, \aj, 115, 1693 
\bibitem[Cox (1981)]{cox81}
	Cox, D.P., \apj, 245, 534
\bibitem[Cui \etal (1996)]{cui96}
	Cui, W., Sanders, W.T., McCammon, D., Snowden, S.L. \&
        Womble, D.S., 1996, \apj, 468, 102
\bibitem[Dahlem \etal (1993)]{dahlem93}
	Dahlem, M., Golla, G., Whiteoak, J. B., Wielebinski, R.,
	 Huettemeister, S. \&  Henkel, C., 1993, \aap, 270, 29
\bibitem[Dahlem \etal (2001)]{dahlem01}
	Dahlem, M., Lazendic, J.S., Haynes, R.F., Ehle, M. \& 
	Lisenfeld, U., 2001, \aap, 374, 42
\bibitem[\protect\citeauthoryear{Dahlem, Lisenfeld \& Golla}{Dahlem \etal}{1995}]{dahlem95}
        Dahlem, M., Lisenfeld, U. \& Golla, G., 1995, \apj, 444, 119
\bibitem[\protect\citeauthoryear{Dahlem, Weaver \& Heckman}{Dahlem \etal}{1998}]{dwh98}
        Dahlem, M., Weaver, K.A. \& Heckman, T.M. 1998,
        \apjs, 118, 401
\bibitem[Dale \etal (2000)]{dale2000}
	Dale, D.A., \etal, 2000, \aj, 120, 583
\bibitem[\protect\citeauthoryear{de Grijs, O'Connell \& Gallagher}{de Grijs \etal}{2001}]{degrijs01}
	de Grijs, R., O'Connell, R.W. \& Gallagher, J.S.,
	2001, \aj, 121, 768
\bibitem[Dekel \& Silk (1986)]{dekel86}
	Dekel A., Silk J., 1986, \apj, 303, 39
\bibitem[Dettmar (1992)]{dettmar92}
	Dettmar, R.-J., 1992, Fundamentals of Cosmic Physics, 15, 143
\bibitem[Dettmar (1998)]{dettmar98}
	Dettmar, R.-J., 1998, in The Local Bubble and Beyond, 
        D., Breitschwerdt, M.J., Freyberg, \& J. Truemper,
	(Springer-Verlag: Berlin Heidelberg New York), 527  
\bibitem[de Vaucouleurs \etal (1991)]{rc3}
	de Vaucouleurs G., de Vaucouleurs A., Corwin Jr. H.G., Buta R.J.,
        Paturel G. \& Fouque P., 1991, {\it Third Reference Catalogue 
	of Bright Galaxies (RC3)} (Springer-Verlag: New York)
\bibitem[De Young \& Heckman (1994)]{deyoung94}
	De Young, D.S. \& Heckman T.M., 1994, \apj, 431, 598
\bibitem[Dickey \& Lockman (1990)]{dickey90}
	Dickey, J.M. \& Lockman, F.J., 1990, \araa. 28, 215. 
\bibitem[Douglas \etal (1996)]{douglas96}
	Douglas, J.N., Bash, F.N., Arakel Bozyan, F.,
        Torrence, G.W., Wolfe, C., 1996, \aj, 111, 1945
\bibitem[Ehle \etal (1998)]{ehle98}
	 Ehle, M., Pietsch, W.,  Beck, R. \& Klein, U., 1998, \aap, 329, 39
\bibitem[Elfhag \etal (1996)]{elfhag96}
	Elfhag, T., Booth, R.S., Hoglund, B.,
	Johansson, L.E.B. \& Sandqvist, Aa., 1996, \aaps, 115, 439
\bibitem[Elmouttie \etal (1997)]{elmouttie97}
	Elmouttie, M., Haynes, R.F., Jones, K.L., Ehle, M.,
	Beck, R., Harnett, J.I. \& Wielebinski, R., 1997, \mnras, 284, 830
\bibitem[Engelbracht \etal (1998)]{engelbracht98}
        Engelbracht, C.W., Rieke, M. J., Rieke, G.H.,
        Kelly, D.M., Achtermann, J.M., 1998, \apj, 505, 639
\bibitem[Fabbiano (1989)]{fabbiano89}
	Fabbiano, G. 1989, \araa, 27, 87
\bibitem[\protect\citeauthoryear{Fabbiano, Heckman \& Keel}{Fabbiano \etal}{1990}]{fhk90}
	Fabbiano, G., Heckman, T.M., Keel, W.C., 1990, \apj, 355, 442
\bibitem[Fabbiano \& Juda (1997)]{fabbiano97}
	Fabbiano, G. \& Juda, J.Z., 1997, \apj, 476, 666
\bibitem[Fabbiano \& Shapley (2002)]{fabbiano02}
	Fabbiano, G. \& Shapley, A., 2002, \apj, 565, 908
\bibitem[\protect\citeauthoryear{Ferrara, Pettini \& Shchekinov}{Ferrara \etal}{2000}]{fps00}
	Ferrara, A., Pettini, M. \& Shchekinov, Y., 2000, \mnras, 319, 539
\bibitem[Ferrara \& Tolstoy (2000)]{ferrara_tolstoy00}
	Ferrara, A. \& Tolstoy, E., 2000, \mnras, 313, 291
\bibitem[Ferri\`ere (2001)]{ferriere01}
	Ferri\`ere, K.M., 2001, Rev.~Mod.~Phys., 73, 1031
\bibitem[Figer \etal (2002)]{figer02}
	Figer, D.F., \etal, 2002, \apj, 581, 258
\bibitem[F\"{o}rster Schreiber \etal (2001)]{forster01}
	F\"{r}ster Schreiber, N.M., Genzel, R., Lutz, D., Kunze, D. 
	\& Sternberg, A., 2001, \apj, 552, 544
\bibitem[Freedman \etal (1994)]{freedman94}
	Freedman, W.L., \etal, 1994, \apj, 427, 628
\bibitem[Garnett (2002)]{garnett02}
	Garnett, D.R., 2002, \apj, 581, 1019
\bibitem[Golla (1999)]{golla99}
	Golla, G., 1999, \aap, 345, 778
\bibitem[Golla \& Wielebinski (1994)]{golla94b}
	Golla, G. \& Wielebinski, R., 1994, \aap, 286, 733
\bibitem[Hameed \& Devereux (1999)]{hameed99}
	Hameed, S. \& Devereux, N., 1999, \aj, 118, 730
\bibitem[Heckman (1999)]{heckman99}
	Heckman, T.M., 1999, \apss, 266, 3
\bibitem[\protect\citeauthoryear{Heckman, Armus \& Miley}{Heckman \etal}{1990}]{ham90}
        Heckman, T. M., Armus, L., \& Miley, G. K.
        1990, \apjs, 74, 833
\bibitem[Heckman \etal (2000)]{heckman00}
	Heckman, T.M., Lehnert, M.D., Strickland, D.K. \& Armus, L.,
	2000, \apjs, 129, 493
\bibitem[Heckman \etal (2001)]{heckman01}
Heckman, T.M., Sembach, K.R., Meurer, G.R., Strickland, D.K., 
        Martin, C.L., Calzetti, D. \& Leitherer, C., 2001, \apj, 554, 1021
\bibitem[Hillenbrand \& Hartmann (1998)]{hillenbrand98}
	Hillenbrand, L.A. \& Hartmann, L.W., 1998, \apj, 492, 540
\bibitem[Hollenbach \& Tielens (1997)]{hollenbach97}
	Hollenbach, D.J. \& Tielens, A.G.G.M., 1997, \araa, 35, 179
\bibitem[Hoopes \etal (2003)]{hoopes03}
	Hoopes, C.G., Heckman, T.M., Strickland, D.K. \& Howk, J.C., 2003,
	\apjl, 596, L175
\bibitem[\protect\citeauthoryear{Hoopes, Walterbos \& Rand}{Hoopes \etal}{1999}]{hoopes99}
	Hoopes, C.G., Walterbos, R.A.M. \& Rand, R.J.,
	1999, \apj, 552, 669
\bibitem[Howk \& Savage (1999)]{howk99}
	Howk, J.C. \& Savage, B.D., 1999, \aj, 117, 2077
\bibitem[Ichikawa \etal (1995)]{ichikawa95}
	Ichikawa, T., Yanagisawa, K., Itoh, N., Tarusawa, K., 
	van Driel, W. \& Ueno, M., 1995, \aj, 109, 2038
\bibitem[Irwin \& Seaquist (1991)]{irwin91}
	Irwin, J.S. \& Seaquist, E.R., 1991, \apj, 371, 111
\bibitem[Irwin \& Sofue (1996)]{irwin96}
	Irwin, J.S. \& Sofue, Y., 1996, \apj, 464, 738
\bibitem[Jarret \etal (2003)]{2mass_largegals}
	Jaaret, T.H., Chester, T., Cutri, R., Schneider, S. \& 
	Huchra, J.P., 2003, \aj, 125, 525
\bibitem[Karachentsev \& Sharina (1997)]{karachentsev97}
	Karachentsev I.D. \& Sharina M.E., 1997, \aap, 324, 457
\bibitem[Katz \etal (2003)]{katz03}
	Katz, N., Keres, D., Dav\'{e}. R. \& Weinberg, D.A., 2003,
	in ``The IGM/Galaxy Connection: The Distribution of Baryons at z=0'',
	Eds. J.L. Rosenberg \& M.E. Putman (Kluwer: Dordrecht), 185
\bibitem[Kennicutt (1998a)]{kennicutt98a}
	Kennicutt, R.C., 1998a, \apj, 498, 541
\bibitem[Kennicutt (1998b)]{kennicutt98b}
	Kennicutt, R.C., 1998b, \araa, 36, 189
\bibitem[Koo \& McKee (1992)]{koo92}
	Koo, B.-C. \& McKee, C.F., 1992, \apj, 388, 93
\bibitem[Koorneef (1993)]{koorneef93}
	Koorneef, J., 1993, \apj, 403, 581
\bibitem[\protect\citeauthoryear{Kroupa, Tout \& Gilmore}{Kroupa \etal}{1993}]{kroupa93}
	Kroupa, P., Tout, C.A., \& Gilmore, G., 1993, \mnras, 262, 545
\bibitem[van der Kruit (1984)]{kruit84}
	Kruit, P.C. van der, 1984, \aap, 140, 470
\bibitem[van der Kruit \& de Grijs (1999)]{vanderkruit99}
	Kruit, P.C. van der, \& de Grijs., R., 1999, \aap, 352, 129
\bibitem[Kuntz \& Snowden (2000)]{kuntz00}
	Kuntz, K.D. \& Snowden, S.L., 2000, \apj, 543, 195
\bibitem[Larson (1974)]{larson74}
	Larson, R.B., 1974, \mnras, 169, 229
\bibitem[Lehnert \& Heckman (1995)]{lehnert95}
	Lehnert, M. \& Heckman, T.M., 1995, \apjs, 97, 89
\bibitem[Lehnert \& Heckman (1996)]{lehnert96}
	Lehnert, M. \& Heckman, T.M., 1996, \apj, 472, 546
\bibitem[Leitherer \& Heckman (1995)]{lh95}
	Leitherer, C. \& Heckman, T.M., 1995, \apjs, 96, 9
\bibitem[\protect\citeauthoryear{Lira, Johnson \& Lawrence}{Lira \etal}{2002}]{lira02}
	Lira, P., Johnson, R. \& Lawrence, A., 2002, submitted to \mnras 
	(astro-ph/0206123) 
\bibitem[Lynds \& Sandage (1963)]{lynds63}
	Lynds, C.R. \& Sandage, A.R., 1963, \apj, 137, 1005
\bibitem[\protect\citeauthoryear{McCarthy, Heckman \& van Bruegel}{McCarthy \etal}{1987}]{mccarthy87}
        McCarthy, P.J., Heckman, T.M. \& van Breugel, W. 1987,
        \aj, 92, 264
\bibitem[McKee (1995)]{mckee95}
	McKee, C.F., 1995, in ASP Conf Series 80, The Physics of the
	Interstellar Medium, A. Ferrara, C.F. McKee, C. Heiles \&
	P.R. Shapiro (San Francisco: ASP), 292
\bibitem[McKee \& Ostriker (1977)]{mckee77}
	McKee, C.F. \& Ostriker, J.P., 1977, \apj, 218, 148 
\bibitem[McKeith \etal (1995)]{mckeith95}
	McKeith, C.D., Greve, A., Downes, D. \& Prada, F., 1995,
	\aap, 293, 703
\bibitem[McLeod \etal (1993)]{mcleod93}
     McLeod, K.K., Rieke, G.H., Rieke, M.J. \& Kelly, D.M.,
	 1993, \apj, 412, 111
\bibitem[Mac Low (1996)]{maclow96}
	Mac Low, M.-M., 1996, in ``The Interplay between Massive Star
	Formation, the ISM, and Galaxy Evolution,'' eds. D. Kunth \etal,
	(Paris: Editions Fronti\`{e}res), 169
\bibitem[Mac Low \& Ferrara (1999)]{maclow99}
	Mac Low, M.-M. \& Ferrara, A., 1999, \apj, 513, 142
\bibitem[Mac Low \& McCray (1988)]{maclow88}
	Mac Low, M.-M. \& McCray, R., 1988, \apj, 324, 776 
\bibitem[\protect\citeauthoryear{Mac Low, McCray \& Norman}{Mac Low \etal}{1989}]{maclow89}
	Mac Low, M.-M., McCray, R., \& Norman, M.L., 1989, \apj, 337, 141 
\bibitem[Malamuth \& Heap (1994)]{malamuth94}
	Malumuth, E.M. \& Heap, S.R., 1994, \aj, 107, 1054
\bibitem[Martin (1999)]{martin_tescape}
	Martin, C.L., 1999, \apj, 513, 156
\bibitem[Martin (1996)]{martin96}
	Martin, C.L., 1996, \apj, 465, 680
\bibitem[Meurer \etal (1995)]{meurer95}
	Meurer, G.R., Heckman, T.M., Leitherer, C., Kinney, A.,
	Robert, C \& Garnett, D.R., 1995, \aj, 110, 2665
\bibitem[\protect\citeauthoryear{Mewe, Kaastra \& Liedahl}{Mewe \etal}{1995}]{mekal}
        Mewe R., Kaastra J. S., 
        Liedahl D. A., 1995, Legacy, 6, 16
\bibitem[Miller \& Veilleux (2003)]{miller03}
        Miller, S.T. \& Veilleux, S., 2003, \apjs, 148, 383
\bibitem[\protect\citeauthoryear{Mori, Ferrara \& Madau}{Mori \etal}{2002}]{mori02}
	Mori, M., Ferrara, A. \& Madau, P., 2002, \apj, 571, 40
\bibitem[\protect\citeauthoryear{Nordgren, Cordes \& Terzian}{Nordgren \etal}{1992}]{nordgren92}
	Nordgren, T.E., Cordes, J.M. \& Terzian, Y., 1992, \aj, 104, 1465
\bibitem[Norman \& Ferrara (1996)]{norman96}
	Norman, C.A. \& Ferrara, A., 1996, \apj, 467, 280
\bibitem[Norman \& Ikeuchi (1989)]{norman89}
	Norman, C. A. \& Ikeuchi, S., 1989, \apj, 345, 372
\bibitem[O'Connell \etal (1995)]{oconnell95}
	O'Connell, R.W., Gallagher, J.S., Hunter, D.A. \& Colley, W.N.,
	1995, \apjl, 446, L1
\bibitem[\protect\citeauthoryear{Oey, King \& Parker}{Oey \etal}{2003}]{oey03}
	Oey, M.S., King, N.L. \& Parker, J.Wm., 2003, \aj, in press
	(astro-ph/0312051)
\bibitem[Olling (1996)]{olling96}
	Olling, R.P., 1996, \aj, 112, 457
\bibitem[Ott \etal (2001)]{ott2001}
	Ott, M., Whiteoak, J.B., Henkel, K. \& Wielebinski, R.,
	2001, \aap, 372, 463
\bibitem[Pence (1981)]{pence81}
	Pence, W.D., 1981, \apj, 247, 473
\bibitem[Press \etal (1992)]{numerical_recipes}
	Press, W.H., Teukolsky, S.A., Vetterling, W.T. \&
	Flannery, B.P., 1992, in Numerical Recipes in FORTRAN: The Art
	of Scientific Computing -- Second Edition (New York: Cambridge
	University Press), 637
\bibitem[Puche \& Carignan (1988)]{puche88}
        Puche, D., Carignan, C., 1988, \aj, 95, 1025
\bibitem[Puche, Carignan \& van Gorkom (1991)]{puche91}
        Puche, D., Carignan, C., van Gorkom, J.H., 1991, \aj, 101, 456
\bibitem[Pudritz \& Fiege (2000)]{pudritz00}
	Pudritz, R.E. \& Fiege, J.D., 2000, in ASP Conf. Series 168,
	New Perspectives on the Interstellar Medium,
	A.R. Taylor, T.L. Landecker \& G. Joncas, (San Francisco: ASP), 235
\bibitem[Rand (1994)]{rand94}
	Rand, R., 1994, \aap, 285, 833
\bibitem[Rand (1996)]{rand96}
	Rand, R.J., 1996, \apj, 462, 712
\bibitem[Read \& Ponman (2001)]{read01}
	Read, A.M. \& Ponman, T.J., 2001, \mnras, 328, 127
\bibitem[\protect\citeauthoryear{Read, Ponman \& Strickland}{Read \etal}{1997}]{rps97}
	Read, A.M., Ponman, T.J., Strickland, D.K., 1997, \mnras, 286, 626
\bibitem[\protect\citeauthoryear{Recchi, Matteucci \& D'Ercole}{Recchi \etal}{2002}]{recchi_vulcano}
	Recchi, S. Matteucci, F. \& D'Ercole, A., 2002, in ``Chemical 
	Enrichment 
        of the ICM and IGM,'' Eds. R. Fusco-Femiano \& F. Matteucci, 
	(ASP: San Francisco), 387
\bibitem[Reynolds (1989)]{reynolds89}
	Reynolds, R.J., 1989, \apjl, 339, L29 
\bibitem[Rice \etal (1988)]{rice88}
	Rice, W., Lonsdale, C.J., Soifer, B.T., Neugebauer, G.,
	Kopan, E.L., Llyod, L.A., deJong, T. \& Habing, H.J.,
	1988, \apjs, 68, 91
\bibitem[Rieke \etal (1993)]{rieke93}
	Rieke, G.H., Loken, K., Rieke, M.J. \& Tamblyn, P.,
	 1993, \apj, 412, 99 
\bibitem[Rossa \& Dettmar (2003)]{rossa03}
	Rossa, J. \& Dettmar, R.-J., 2003, \aap, 406, 493
\bibitem[Rossa \& Dettmar (2000)]{rossa00}
	Rossa, J. \& Dettmar, R.-J., 2000, \aap, 359, 433
\bibitem[\protect\citeauthoryear{Roth, Mould \& Davies}{Roth \etal}{1991}]{roth91}
	Roth, J., Mould, J.R. \& Davies, R.D., 1991, \aj, 102, 1303
\bibitem[Rupen (1991)]{rupen91}
	Rupen, M.P., 1991, \aj, 102, 48
\bibitem[Sanders \& Mirabel (1996)]{sander96}
	Sanders, D.B. \& Mirabel, I.F., 1996, \araa, 34, 749
\bibitem[Satyapal \etal (1997)]{satyapal97}
	Satyapal, S., Watson, D.M., Pipher, J.L., Forrest, W.J.,
	Greenhouse, M.A., Smith, H.A., Fischer, J. \& Woodward, C.E.,
	1997, \apj, 483, 148
\bibitem[Sembach \& Danks (1994)]{sembach94}
	Sembach, K.R. \& Danks, A.C., 1994, \aap, 289, 539
\bibitem[Shapiro \& Field (1976)]{shapiro76}
	Shapiro, P.R. \& Field, G.B., 1976, \apj, 205, 762
\bibitem[\protect\citeauthoryear{Shapiro, Giroux \& Babul}{Shapiro \etal}{1994}]{shapiro94}
	Shapiro, P.R., Giroux, M.L. \& Babul, A., 1994,  \apj, 427, 25
\bibitem[\protect\citeauthoryear{Shapley, Fabbiano \& Eskridge}{Shapley \etal}{2001}]{shapley01}
	Shapley, A., Fabbiano, G. \& Eskridge, P., 2001, \apjs, 137, 139
\bibitem[Shopbell \& Bland-Hawthorn (1998)]{shopbell98}
	Shopbell, P.L., Bland-Hawthorn, J., 1998, \apj, 493, 129 
\bibitem[Shull \& Slavin (1994)]{shull_halo}
	Shull, J.M. \& Slavin, J.D., 1994, \apj, 427, 784
\bibitem[Silich \& Tenorio-Tagle (1998)]{silich98}
	Silich, S.A. \& Tenorio-Tagle, G., 1998, \mnras, 299, 249
\bibitem[Silich \& Tenorio-Tagle (2001)]{silich01}
	Silich, S.A. \& Tenorio-Tagle, G., 2001, \apj, 552, 91
\bibitem[Snowden \& Pietsch (1995)]{snowden95}
	Snowden, S.L. \& Pietsch, W., 1995, \apj, 452, 627
\bibitem[Sofue (1997)]{sofue97}
	Sofue, Y., 1997, \pasj, 49, 17
\bibitem[Soifer \etal (1989)]{soifer89}
	Soifer, B.T., \etal, 1989, \aj, 98, 1766
\bibitem[Soifer \etal (1987)]{soifer87}
	Soifer, B.T., Sanders, D.B., Madore, B.F., Neugebauer, G.,
	Danielson, G.E., Elias, J.H., Lonsdale, C.J. \&
	Rice, W.L., 1987, \apj, 320, 238
\bibitem[Sommer-Larsen \etal (2003)]{sommerlarsen02}
	Sommer-Larsen, J., Toft, S., Rasmussen, J., Pedersen, K.,
	Gotz, M. \& Portinari, L., 2003, \apss, 284, 693
\bibitem[Sorai \etal (2000)]{sorai2000}
	Sorai, K., Nakai, N., Kuno, N., Nishiyama, K. \&
	Hasegawa, T., 2000, \pasj, 52, 785
\bibitem[Strickland (2002)]{strickland_vulcano}
	Strickland, D.K., 2002, in ``Chemical 
	Enrichment 
        of the ICM and IGM,'' Eds. R. Fusco-Femiano \& F. Matteucci, 
	(ASP: San Francisco), 387
\bibitem[Strickland \etal (2003)]{strickland_iau}
	Strickland, D.K., Heckman, T.M., Colbert, E.J.M., Hoopes, C.G.
	\& Weaver, K.A., 2003, in 
	``A Massive Star Odyssey, from Main Sequence to Supernova,''
	Eds. K.A. van der Hucht, A. Herrero \& C. Esteban 
	(ASP: San Francisco), 612
\bibitem[Strickland \etal (2004)]{strickland03}
	Strickland, D.K., Heckman, T.M., Colbert, E.J.M., Hoopes, C.G.
	\& Weaver, K.A., 2003, \apjs, in press (Paper I)
\bibitem[Strickland \etal (2002)]{strickland02}
	Strickland, D.K., Heckman, T.M., Weaver, K.A., 
	Hoopes, C.G. \& Dahlem, M., 2002, \apj, 568, 689
\bibitem[Strickland \etal (2000)]{strickland00}
	Strickland, D.K., Heckman, T.M., Weaver, K.A. \& Dahlem, M.,
	2000, \aj, 120, 2965
\bibitem[Strickland \& Stevens (1998)]{ss98}
	Strickland, D.K. \& Stevens, I.R., 1998, \mnras, 297, 747
\bibitem[Strickland \& Stevens (1999)]{ss99}
	Strickland, D.K. \& Stevens, I.R., 1999, \mnras, 306, 43
\bibitem[Strickland \& Stevens (2000)]{ss2000}
	Strickland, D.K. \& Stevens, I.R., 2000, \mnras, 314, 511

\bibitem[Suchkov \etal (1994)]{suchkov94}
        Suchkov, A.A., Balsara, D,S.,
        Heckman, T.M., Leitherer, C., 1994, \apj, 430, 511
\bibitem[Tenorio-Tagle (1996)]{tt96}
	Tenorio-Tagle, G., 1996, \aj, 111, 1641
\bibitem[Thornton \etal (1998)]{thornton98}
	Thornton, K., Gaudlitz, M., Janka, H.-Th. \& Steinmetz, M.,
        1998, \apj, 500, 95
\bibitem[Toft \etal (2002)]{toft02}
	Toft, S., Rasmussen, J., Sommer-Larsen, J. \& Pedersen, K.,
	2002, \mnras, 335, 799
\bibitem[Tomisaka \& Bregman (1993)]{tomisaka93}
	Tomisaka, K., \& Bregman, J.N., 1993, \pasj, 45, 513
\bibitem[Tremonti \etal (2001)]{tremonti01}
	Tremonti, C., Calzetti, D., Leitherer, C. \& Heckman, T.M.,
	2001, \apj, 555, 322
\bibitem[Tremonti \etal (in preparation)]{tremonti}
	Tremonti, C.A., \etal, in preparation
\bibitem[Tully \& Fouqu\'{e} (1985)]{tully85}
	Tully, R.B. \& Fouqu\'{e}, P., 1985, \apjs, 58, 67
\bibitem[\protect\citeauthoryear{Vogler, Pietsch \& Kahabka}{Vogler \etal}{1995}]{vogler95}
	Vogler, A., Pietsch, W. \& Kahabka, P., 1995, \aap, 305, 71
\bibitem[Voit (1996)]{voit96}
	Voit, G.M., 1996, \apj, 465, 548
\bibitem[\protect\citeauthoryear{Wainscoat, de Jong \& Wesselius}{Wainscoat \etal}{1987}]{wainscoat87}
	Wainscoat, R.J., de Jong, T. \& Wesselius, P.R., 1987, \aap, 181, 225
\bibitem[Wang \etal (2001)]{wang2001}
	Wang, Q.D., Immler, S., Walterbos, R., Lauroesch, J.T. \&
	Breitschwerdt, D., 2001, \apjl, 555, L99
\bibitem[Wang \etal (1995)]{wang95}
	Wang, Q.D., Walterbos, R.A.M., Steakley, M.F.,
	Norman, C.A. \& Braun, R., 1995, \apj, 439, 176
\bibitem[Weaver \etal (1977)]{weaver77}
        Weaver R., McCray R., Castor J., Shapiro P., Moore R.,
        1977, \apj, 218, 377
\bibitem[Weliachew, Fomalont \& Greisen (1984)]{weliachew84}
	Weliachew, I., Fomalont, E.B. \& Greisen, E.W.,
	1984, \aap, 137, 335
\bibitem[Wiseman \& Ho (1998)]{wiseman98}
	Wiseman, J.J. \& Ho, P.T.P., 1998, \apj, 502, 676
\bibitem[Wolfire \etal (2003)]{wolfire03}
	Wolfire, M.G., McKee, C.F., Hollenbach, D. \& Tielens, A.G.G.M.,
	2003, \apj, 587, 278
\end{thebibliography}
\end{document}